\documentclass[galaxies,review,accept,pdftex,moreauthors]{Definitions/mdpi} 

\firstpage{1} 
\pubvolume{1}
\issuenum{1}
\articlenumber{0}
\pubyear{2025}
\copyrightyear{2025}
\externaleditor{~}
\datereceived{13 March 2025} 
\daterevised{12 April 2025} 
\dateaccepted{15 April 2025} 
\datepublished{ } 
\hreflink{https://doi.org/} 

\usepackage{aas_macros}

\usepackage{todonotes}

\definecolor{mygreen}{RGB}{0,128,0}

\graphicspath{{pic/}}

\Title{Betelgeuse, the Prototypical Red Supergiant}
\TitleCitation{Betelgeuse, the Prototypical Red Supergiant}

\Author{Andrea K. Dupree $^{1,}$*$^{,\dagger}$\orcidA{} 
  and Miguel Montargès $^{2,}$*$^{,\dagger}$\orcidB{}}

\AuthorNames{Andrea K. Dupree and Miguel Montargès}

\AuthorCitation{Dupree, A.K.; Montargès, M.}

\address{%
$^{1}$ \quad Center for Astrophysics | Harvard and Smithsonian, Cambridge, MA, 02138 USA\\
$^{2}$ \quad LIRA, Observatoire de Paris, Université PSL, Sorbonne Université, Université Paris Cité, CY Cergy Paris Université, CNRS, 92195 Meudon CEDEX, France}

\corres{Correspondence: adupree@cfa.harvard.edu (A.K.D.); miguel.montarges@obspm.fr (M.M.)}

\firstnote{These authors contributed equally to this work.}

\abstract{The behavior of the bright red supergiant, Betelgeuse, is described with results principally from the past 6 years. The review includes  imaging, photometry, and spectroscopy to record the Great Dimming of 2019--2020. This event was followed by a slow ongoing recovery from the massive surface mass ejection after which the stellar characteristics changed. Theoretical simulations address  the cause of this episodic mass ejection and the optical Dimming. Recent publications evaluating the perplexing 2100 day periodicity in the star's brightness and radial velocity provide evidence that Betelgeuse may harbor a companion object. Current attempts at direct detection of this companion are discussed. Betelgeuse provides a well-studied and meaningful example for supergiant stars in our Galaxy and others.}


\begin{document}


\section{Introduction}  

The bright star Betelgeuse has been recognized by humans for many centuries. 
A paleolithic carving  from $\sim$40,000 years ago in the  Geissenkl{\"aosterle Cave appears to replicate the Orion
constellation~\cite{2003UppOR..59...51R}.  Drawings of  Betelgeuse in the constellation 
appear  on the wall of the Lascaux Cave in southeastern France.  These
are  believed to have been created about 21,000 years ago.  The~brightness
of Betelgeuse was  recorded
in pre-telescopic times in Greek and Latin manuscripts,  writings in Rome and China 
over 2000 years ago, and~Greek vases from the V century BC~\cite{2022MNRAS.516..693N}.
Additionally, just over a century ago, the diameter of Betelgeuse  was first measured with an interferometer constructed at
Mt. Wilson Observatory with a 20 foot rigid beam mounted on the
100-inch telescope~\cite{1921ApJ....53..249M}.  Their result was reported in a headline about
a gigantic star---`a colossus of the skies' \cite{NYtimes1920}.  A~workshop proceeding~\cite{2013EAS....60.....K} gives
a broad overview of the star.  Betelgeuse occupies a pivotal place in stellar evolution---helium burning in preparation
for its next stage---a supernova.  Although~over a century of observations exist,  many of its characteristics remain 
puzzling and, as of yet, unexplained.

Supergiant stars play a major role in affecting the interstellar medium with their strong stellar winds,
and enrichment of chemical elements, as~other reviews in this volume have discussed. The~red supergiant (RSG)
star Betelgeuse ($\alpha$ Orionis, HD 39801), classified as  M1-M2 Ia-Iab~\cite{1989ApJS...71..245K},                                                                                                                                                                                                                                                                                                                                                  presents today's  astronomers
as an object giving unique  insight into the  astrophysics of evolved stars.  It is the ninth brightest
star visible in the sky~\cite{1987ADCBu...1..285H}, and~as a cool supergiant, it is
physically the largest among this bright  group. Moreover, Betelgeuse is relatively close
by, thus enabling direct imaging of its surface and extended atmosphere (Figure~\ref{Fig:Betelgeuse_scales}). 
It was  the first stellar surface to be imaged directly~\cite{1996ApJ...463L..29G} except for our own Sun, a much smaller and closer dwarf star. The~ultraviolet image revealed a large hot spot believed to result from 
photospheric  convection.   Subsequent spatially resolved images and interferometric results in the ultraviolet, near~infrared,
millimeter and centimeter wavelengths confirm the presence and variability of convective regions~\cite{2009A&A...503..183O,2009A&A...508..923H,2011A&A...529A.163O,2013EAS....60...77D,2016A&A...588A.130M,2017A&A...602L..10O}. Recently, time monitoring of the photosphere has become possible thanks to an innovative technique: the  linear polarimetry signal in several atomic lines has been interpreted as depolarization of the continuum, allowing map reconstructions of the brightness inhomogeneities on a sphere~\cite{2022A&A...661A..91L,2024A&A...691A.297P}. Thanks to the rapid acquisition cadence of the spectropolarimetric observations, time monitoring of what is interpreted as the convective pattern has become possible, and~theoretical simulations~\cite{2010A&A...515A..12C}
support this interpretation.   All of these characteristics
have made Betelgeuse an attractive star for quantitative study for
over a~century.
		
\begin{figure}[H]
	\includegraphics[width=\textwidth]{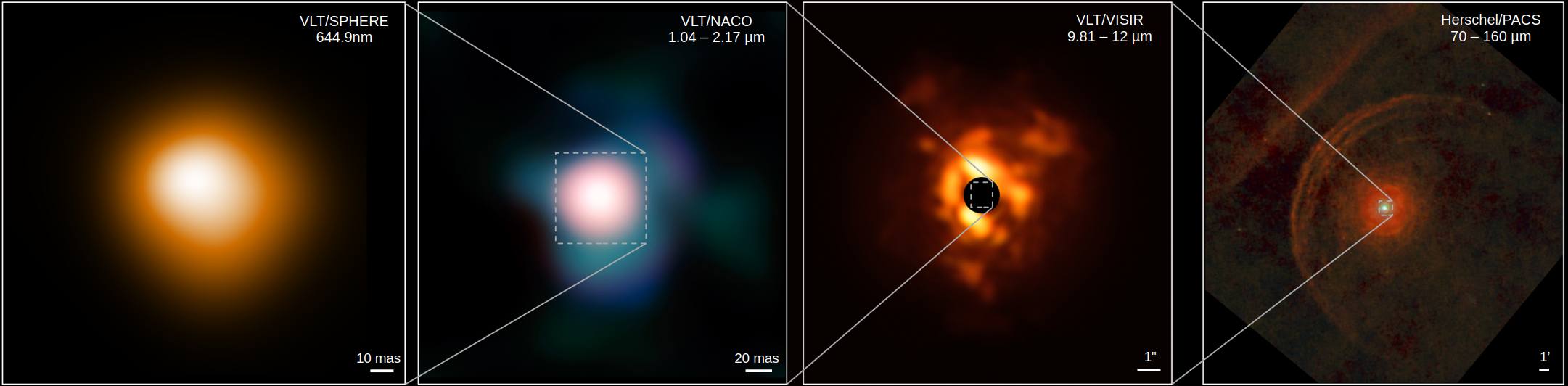}
	\caption{Various spatially resolved observations of Betelgeuse. On~each image North is up and East to the left. The~spatial scale is indicated with a ruler at the bottom right corner of each image. Image inspired by presentations from P.~Kervella. (\textbf{First image}) VLT/SPHERE adaptive optics image from January 2019 at 644.9~nm~\cite{2021Natur.594..365M}. (\textbf{Second image}) VLT/NACO composite image from 1.04 to 2.17~$\upmu$m obtained in January 2009.~\cite{2009A&A...504..115K}. (\textbf{Third image}) VLT/VISIR composite image from December 2019, between 9.81 and 12~$\upmu$m (\url{https://www.eso.org/public/images/eso2003d/}, accessed on 21 April 2025). (\textbf{Fourth image}) Herschel/PACS observations of September 2010 and March 2012 between 70 and 160~$\upmu$m. ESA/Herschel/PACS.~\cite{2012A&A...548A.113D}. \label{Fig:Betelgeuse_scales}}
\end{figure}
	
A  review of Betelgeuse observations and theoretical interpretation
was given in 2023~\cite{2023A&G....64.3.11W}.  Our review begins with
a summary of the physical properties of Betelgeuse and then 
picks up the story beginning in 2019, with~the Great Dimming.  
The road to recovery of the supergiant is followed, concluding with most recent results, and challenges for future~study.

\section{Stellar~Parameters}
\unskip

\subsection{Ultraviolet Surface~Images}

In 1995, the~first direct image of the surface of a star other than the Sun was
made~\cite{1996ApJ...463L..29G}  with the Hubble Space Telescope Faint Object Camera (FOC) in two broad
ultraviolet bands centered at 253 nm and 278 nm (Figure~\ref{Fig:Betelgeuse_rotation}).
The FOC offered good
spatial sampling with pixels of 14.35~mas in the near-UV point spread function.
In the ultraviolet continuum, the~Betelgeuse diameter of $125 \pm 5$~mas at 250~nm exceeds the
optical diameter ($\sim$42~mas,~\cite{2021Natur.594..365M}) by a factor of $\sim$3, thus giving about nine
resolution elements across the stellar disk.  The~Mg II line emission extends even
further reaching a diameter of  about 270~mas or about six stellar radii~\cite{1998AJ....116.2501U,2020ApJ...899...68D}. The~image contained one unresolved
bright area in the southwest quadrant with a temperature difference in
excess of 200~K  on the ultraviolet disk. Surface features were suggested
from  optical interferometric measures (e.g.,~\cite{1997MNRAS.291..819W}).  Such
features were perhaps not unexpected since the appearance of convective
elements on the surface---in the  case of a 15~M$_{\odot}$ star,  90 elements---had been hypothesized previously~\cite{1975ApJ...195..137S}.  However, since convective
heat transport vanishes at the photosphere, where the transport of energy
is taken over by radiation,   it appears likely that chromospheric hot regions
may be associated with shock waves induced in the stellar photosphere~\cite{1998AJ....116.2501U}.
Tomographic techniques have recently demonstrated the development of these
shocks in the Betelgeuse photosphere~\mbox{\cite{2021A&A...650L..17K,2024A&A...685A.124J}}.  They may
extend to chromospheric levels, perhaps guided by magnetic~fields.

\vspace{-12pt}
\begin{figure}[H]

	\includegraphics[width=.43\textwidth]{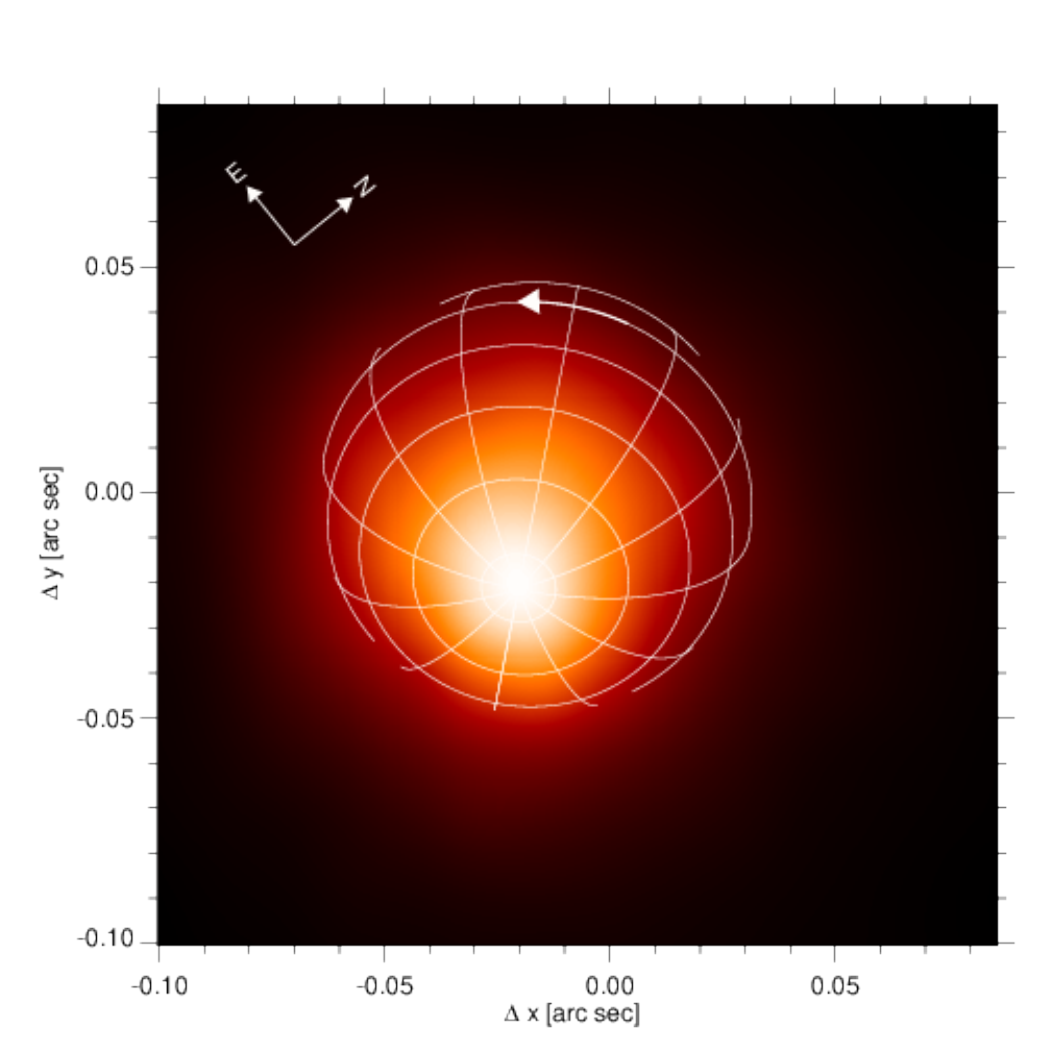}~
	\includegraphics[width=.4\textwidth]{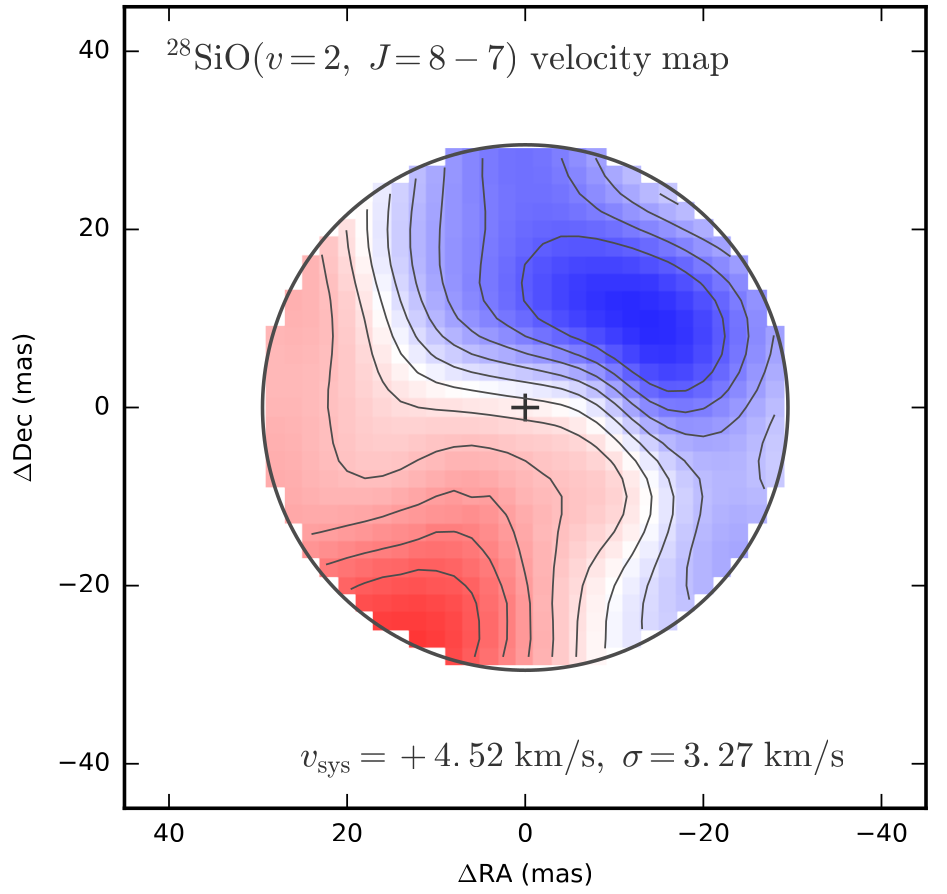}
	\caption{The rotating Betelgeuse. (\textbf{Left}) UV continuum direct  image taken with HST/FOC in March 1995~\cite{1996ApJ...463L..29G} showing the hot spot and  the direction of rotation of the star~\cite{1998AJ....116.2501U}. (\textbf{Right}) Velocity map of Betelgeuse in the $^{28}$SiO~($v = 2$, $J = 8 - 7$) emission line measured over the equivalent continuum disk of the star in 2015--2016~\cite{Kervella2018}. \label{Fig:Betelgeuse_rotation}}
\end{figure}

Following the first images, spectra were obtained with the Small Science Aperture
(220~mas squared) of the Goddard High Resolution Spectrograph (GHRS) on HST.  Different
offset pointings with rapid spatial stepping were made in a perpendicular pattern
across the disk and beyond.  The~changing radial velocity of the Mg II lines revealed
the rotation of the chromosphere shown in Figure~\ref{Fig:Betelgeuse_rotation}.   The~results from
the ultraviolet are complemented by the subsequent measures of the SiO mm emission
from ALMA~\cite{Kervella2018} and demonstrate that the chromosphere is co-rotating with the star. This 
later interferometric image with angular resolution of 18 mas detected a bright spot in the northeast quadrant of the star.  The~results from~\cite{1998AJ....116.2501U,Kervella2018} might suggest that the
polar regions are preferable for the appearance of bright~spots.

However, additional images with the FOC (Figure~\ref{Fig:Betegeuse_UV_2013}) were obtained over 4 years~\cite{2013EAS....60...77D}. The~images illustrate the changing strength and position of the bright areas as seen in the ultraviolet.
Spatially resolved ultraviolet spectra of several singly and doubly ionized species have revealed non-radial chromospheric motions that can flow in opposite directions with velocities of $\sim $2~km~s$^{-1}$ \cite{2001ApJ...558..815L}. The~association or not with bright regions is~unknown.

\begin{figure}[H]
	
	\includegraphics[width=.7\textwidth]{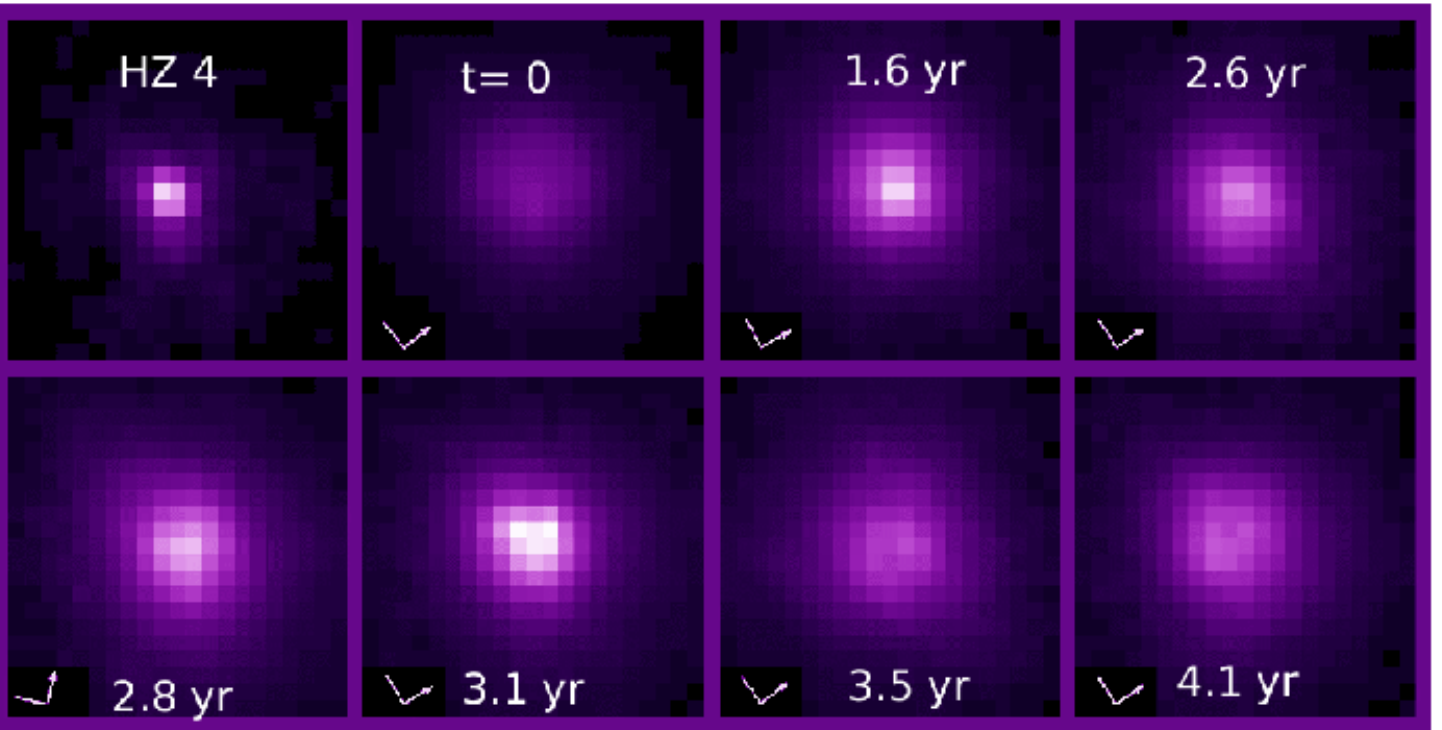}
	\includegraphics[width=.7\textwidth]{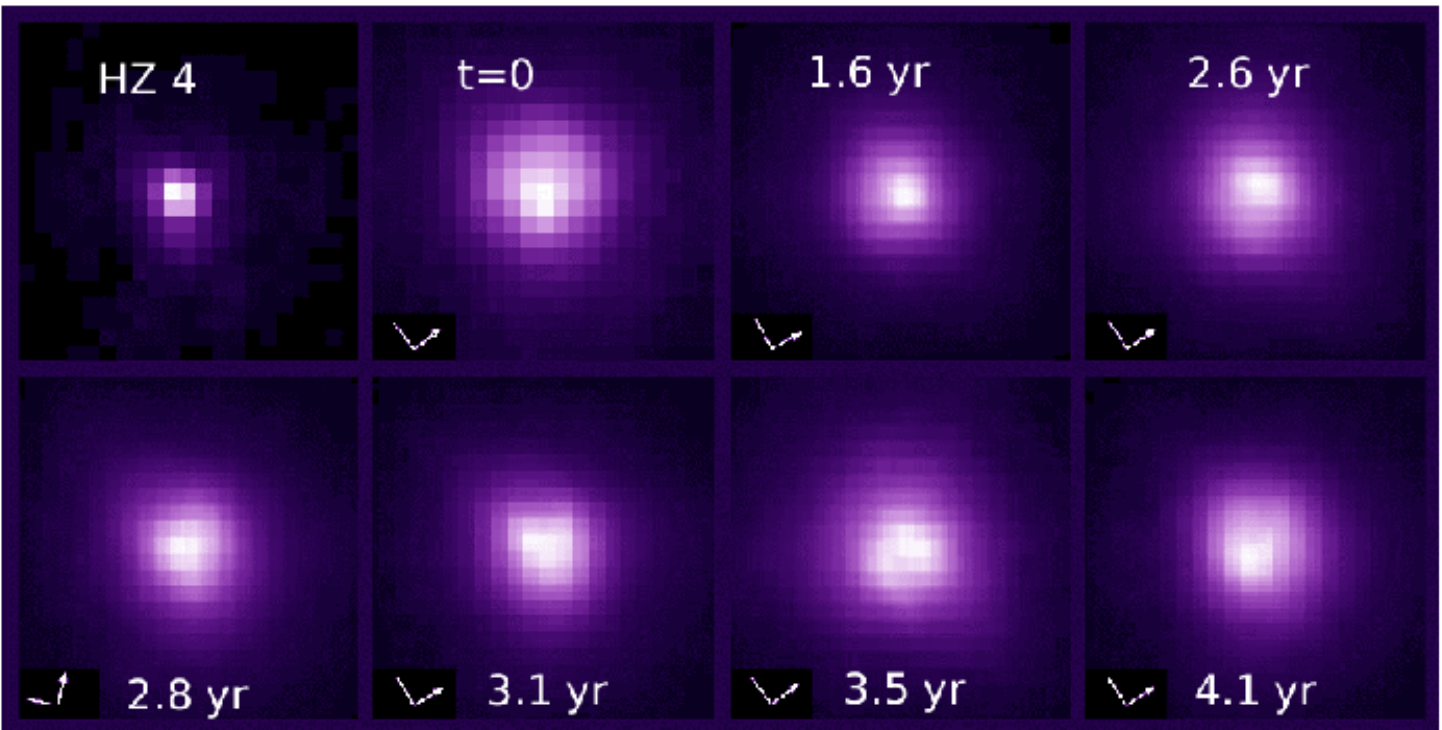}
	\caption{Betelgeuse images in the 253~nm continuum taken with the HST/FOC beginning in 1995 March and extending to 1999 March. The~star HZ 4 is a single white dwarf which is a point source in diameter as compared to Betelgeuse. ({\bf Top panel}) Images scaled to the same exposure time (3559~s) illustrating the strong variation in the ultraviolet flux. ({\bf Lower panel}) Images scaled to the brightest pixel in the image which demonstrates the different location and extent of the bright area in the chromosphere~\cite{2013EAS....60...77D}.\label{Fig:Betegeuse_UV_2013}}
\end{figure}

\subsection{The Distance~Tension}

Betelgeuse is the second closest RSG (after Antares, $\alpha$~Sco), and~as such its angular diameter makes it among the largest observable stars from Earth. Ironically, these characteristics make it very difficult to have a proper estimate of its physical parameters. Most of those depend on a reliable and precise value for its distance. Unfortunately, the~brightness of Betelgeuse prevents any observation with \textit{Gaia}. Even if such observations were possible, the~problem would not be solved. \textit{Gaia} measures the position of the photocenter of a star at several epochs to derive a parallax value. In~Betelgeuse's case, the~parallax is an order of magnitude smaller than its angular diameter. The powerful convective motion at its surface can cause a larger photocenter displacement than its parallax~\cite{2011A&A...528A.120C,2022A&A...661L...1C}. The~initial modern estimate of Betelgeuse's distance was obtained with \textit{Hipparcos}: $131^{+35}_{-23}$~pc~\cite{1997ESASP1200.....E,1997A&A...323L..49P}. A~decade later, the~revised \textit{Hipparcos} astrometry proposed a distance of $152 \pm 20$~pc, with~a "Cosmic Noise" of 2.4~mas~\cite{2007A&A...474..653V,2017AJ....154...11H}. Simultaneously, a~combination of Very Large Array (VLA) radio position and \textit{Hipparcos} Intermediate Astrometric Data (IAD) gave a distance of $197^{+55}_{-35}$~pc~\cite{2008AJ....135.1430H}. Sub-millimeter and millimeter observations are probing a region above the convective photosphere and below the hot chromosphere that is  sensitive to the brightness fluctuations caused by temperature inhomogeneities. The~last such parallax measurement combining the revised \textit{Hipparcos} IAD, and~both VLA and e-MERLIN monitoring of the proper motion of the star between 1982 and 2016 has produced a value of $222^{+48}_{-34}$~pc~\cite{2017AJ....154...11H}. Betelgeuse never ceases to become more distant and bigger since its measured angular diameter remains constant. More recently, another estimate has been provided by identifying the first overtone pulsation mode within the light curve of the star, and~comparing it with models from the Modules for Experiments in Stellar Astrophysics (MESA), to~determine the stellar radius and luminosity of Betelgeuse. This leads to a distance of $168^{+27}_{-15}$~pc~\cite{2020ApJ...902...63J}. The~broad ranges of values derived for the various  stellar parameters are related to this uncertainty on the~distance. 

In Table~\ref{Tab:Stellar_Param}, we summarize the available parameters, and~attach them to a distance solution when~relevant.

\begin{table}[H] 
	\caption{Fundamental parameters of Betelgeuse, partly compiled from~\cite{2022ApJ...936...18D}. \label{Tab:Stellar_Param}}
	
\begin{adjustwidth}{-\extralength}{0cm}

	\begin{tabularx}{\fulllength}{CCcC}
		\toprule
		\textbf{Property}	& \textbf{Value}	& \textbf{Reference} & \textbf{Remarks}\\
		\midrule
		Temperature & $3650 \pm 50$ K & \cite{2005ApJ...628..973L} & Spectrometry and MARCS~models \\
		Spectral type & M1-M2Ia-Iab & \cite{1989ApJS...71..245K} & Photographic~spectra \\
		Radial velocity & $+21.91 \pm 0.51$ km s$^{-1}$ & \cite{Famaey2005} & Integrated photospheric~spectrum \\
		Rotation velocity & $v_\mathrm{eq} \sin i = 5.47 \pm 0.25$~km~s$^{-1}$ & \cite{Kervella2018} & Equatorial velocity, $^{28}$SiO ($v = 3 - 2$, $J = 8 - 7$) \\
		Rotation velocity & None? & \cite{2024ApJ...962L..36M} & RHD~simulations \\
		Chromospheric rotation velocity & 14.6 km s$^{-1}$ & \cite{1998AJ....116.2501U} & UV~spectroscopy \\
		Uniform disk angular diameter---Photosphere & $42.61 \pm 0.05$~mas $42.11 \pm 0.05$~mas & \cite{2021Natur.594..365M} & K band continuum interferometry    January~2019 + February~2020 \\
		Uniform disk angular diameter---Chromosphere & $125 \pm 5$~mas & \cite{1996ApJ...463L..29G} & 250 nm{} continuum~imaging\\
		Distance & $222^{+48}_{-34}$~pc &  \cite{2017AJ....154...11H} & \textit{Hipparcos} IAD + VLA + e-MERLIN\\
		Distance & $168^{+27}_{-15}$~pc & \cite{2020ApJ...902...63J} & Seismic~analysis \\
		Photospheric radius & $1010^{+216}_{-152}$ R$_\odot$ & This work & VLTI January~2019 + \textit{Hipparcos} IAD + VLA + e-MERLIN \\
		Photospheric radius & $764^{+116}_{-62}$ R$_\odot$ & \cite{2020ApJ...902...63J} &  Seismic~analysis \\
		Mass-loss rate & $2.1 \times 10^{-7}$ M$_\odot$ yr$^{-1}$ & \cite{DeBeck2010} & d = 131~pc, CO line~profile \\
		Initial mass & $20^{+5}_{-3}$ M$_\odot$ & \cite{2016ApJ...819....7D} & d $=197 \pm 45$~pc \\
		Current mass & $19.4 - 19.7$ & \cite{2016ApJ...819....7D} & d $=197 \pm 45$~pc \\
		Age & $8.0 - 8.5$ Myr & \cite{2016ApJ...819....7D} & d $=197 \pm 45$~pc \\
		Initial mass & $18 - 21$ M$_\odot$ & \cite{2020ApJ...902...63J} & Seismic~analysis\\
		Current mass & $16.6 - 19$ M$_\odot$ & \cite{2020ApJ...902...63J} & Seismic~analysis\\
		Age & $7 - 11$ Myr & \cite{2020ApJ...902...63J} & Seismic~analysis \\
		Initial mass & $19$ M$_\odot$ & \cite{2023MNRAS.526.2765S} & d = $222^{+48}_{-34}$~pc Seismic~analysis\\
		Current mass & $11 - 12$ M$_\odot$ & \cite{2023MNRAS.526.2765S} & d = $222^{+48}_{-34}$~pc Seismic~analysis \\
		\bottomrule
	\end{tabularx}
\end{adjustwidth}
\end{table}
\unskip

\subsection{The Controversial Rotation~Velocity \label{SubSect:rotation}}

Among other parameters, the~rotation of Betelgeuse has been evaluated several times. In~1998, an~initial measurement was obtained from Hubble Space Telescope (HST) observations with the FOC (Figure~\ref{Fig:Betelgeuse_rotation}, \cite{1998AJ....116.2501U}). This corresponds to the velocity measured for the chromosphere estimated at 2.25~R$_\star$ by the authors. They found a value of 14.6~km~s$^{-1}$.

In 2018, ALMA observations of the $^{28}$SiO ($v = 3 - 2$, $J = 8 - 7$) line showed a velocity field close to solid rotation, along the same direction (North East to South West) (Figure~\ref{Fig:Betelgeuse_rotation}, \cite{Kervella2018}). The~derived equatorial velocity $v_\mathrm{eq} \sin i = 5.47 \pm 0.25$~km~s$^{-1}$ at $R_\mathrm{ALMA} = 29.50 \pm 0.14$~mas (1.4 times the near infrared radius) is quite high. This led several authors to consider that Betelgeuse might have been the result of a past merger~\cite{1998AJ....116.2501U,2017MNRAS.465.2654W,2020ApJ...896...50C,2020ApJ...905..128S}. This scenario would both explain the high rotation velocity of the star, and~its unusual apparent single state which challenges massive star formation processes~\cite{2012Sci...337..444S,2022A&A...663A..26B}.

In 2024,  considerations of radiative hydrodynamics (RHD) simulations suggested that the measured velocity field could have been caused in fact by the convective activity on the photosphere, blurred by the ALMA beam~\cite{2024ApJ...962L..36M}, and~different motions amounting to \mbox{2 km s$^{-1}$} have been observed in the low chromosphere over a period of 15 months~\cite{2001ApJ...558..815L}. This highlights the importance of the characterization of the velocity field of the photosphere of the star: both to understand the convective processes on the surface (and their role in the mass-loss onset mechanism), and~to bring constraints on the evolutionary scenario for Betelgeuse.  Another ALMA measurement would be useful~\cite{2024ApJ...962L..36M} to confirm the signs (or absence) of~rotation.

\subsection{A Dying Star Hiding Its~Age}

How old is Betelgeuse? Or more precisely, how close is it to becoming a supernova (SN), that will be 10\% of the brightness of the full moon according to theoretical supernova models, or~possibly half as bright if it is particularly  luminous? Such an event could last for several weeks in  our sky~\cite{2020RNAAS...4...35G}. In~2016, a~study proposed that Betelgeuse recently entered the helium core burning stage~\cite{2016ApJ...819....7D}. Later, considerations of the evolution of the color of Betelgeuse at the beginning of our era  confirm this scenario~\cite{2022MNRAS.516..693N}. The~seismic analysis that re-evaluated the fundamental stellar parameters found similar results~\cite{2020ApJ...902...63J}.

One study proposed an alternate scenario where Betelgeuse would be in the carbon burning stage, meaning that the SN would be less than $\sim$300~years in the future~\cite{2023MNRAS.526.2765S}. This is based on a seismic analysis and a new estimate of the stellar radius of 1300~R$_\odot$, needed to reinterpret the frequency power spectrum. This exciting scenario stretches the angular diameter measurements to the extreme---for this interpretation is in tension with existing observational constraints~\cite{2023RNAAS...7..119M}. Such a stellar radius for the photosphere would correspond to an angular diameter $> 54$~mas, even at the largest distance estimate ($222^{+48}_{-34}$~pc,~\cite{2017AJ....154...11H}). Such values have only been obtained for the chromosphere and the extended molecular or dusty envelope, but~never for the infrared photosphere~\cite{2014A&A...572A..17M}. Another point raised by such a scenario is the difference between the estimated current mass ($11 - 12$~M$_\odot$) and the zero-age main sequence (ZAMS) mass (19~M$_\odot$). The~missing 7 to 8~M$_\odot$ is expected to have been lost during the RSG stage~\cite{2012A&A...537A.146E}, and~to now be part of the circumstellar environment. However, the~simple fact that Betelgeuse remains brightly visible to the naked eye indicates that its envelope contains little material (contrary to, e.g.,~VY~CMa,~\cite{2021AJ....161...98H}). This is confirmed by observations that indicate that only a fraction of a solar mass is surrounding Betelgeuse: 0.042--0.096~M$_\odot$ \cite{1993ApJS...86..517Y,1997AJ....114..837N,2012MNRAS.422.3433L}.

In any case, considerations of the age of Betelgeuse are necessarily hindered by the uncertainty in its distance, and~even more by considerations of its possible binarity evolution. If~indeed Betelgeuse is the product of a merger (see Section~\ref{SubSect:rotation}), it may have  experienced complex physical and chemical processes, making an age estimate~difficult.

\section{The Great~Dimming}

Betelgeuse has been observed visually, photometrically, and~spectroscopically  for more than a century.
In both optical light and radial velocity, the~star exhibits periodic variability on two time scales: $\sim$400 days and
a long secondary period  (LSP) of about 2100 days (Figure~\ref{Fig:AAVSO_V}).  The~short period is attributed to the fundamental pulsation mode
of the star~\cite{2020ApJ...902...63J}.  The~causes of the long secondary period (LSP) are not understood. Many suggestions have 
been offered as to its origin~\cite{2004ASPC..310..322W}.  These
include  a cycle of giant convective cells, non-radial pulsations,  strange modes, binarity, and~many others which are  discussed in Section~\ref{Sect:LSP},  
and were evaluated in detail~\cite{2024ApJ...977...35G}.   

As can be seen in Figure~\ref{Fig:AAVSO_V}, after~a well-defined $\sim$400-day variation, the~visual magnitude became  
fainter in the last months of 2019 ($\sim$ JD 2458800).  This rapidly
led to an historic optical Dimming in February 2020.  Spectroscopy in the optical and ultraviolet has 
provided insight into the processes occurring in the stellar~atmosphere.

\begin{figure}[H]

	\includegraphics[width=.95\textwidth]{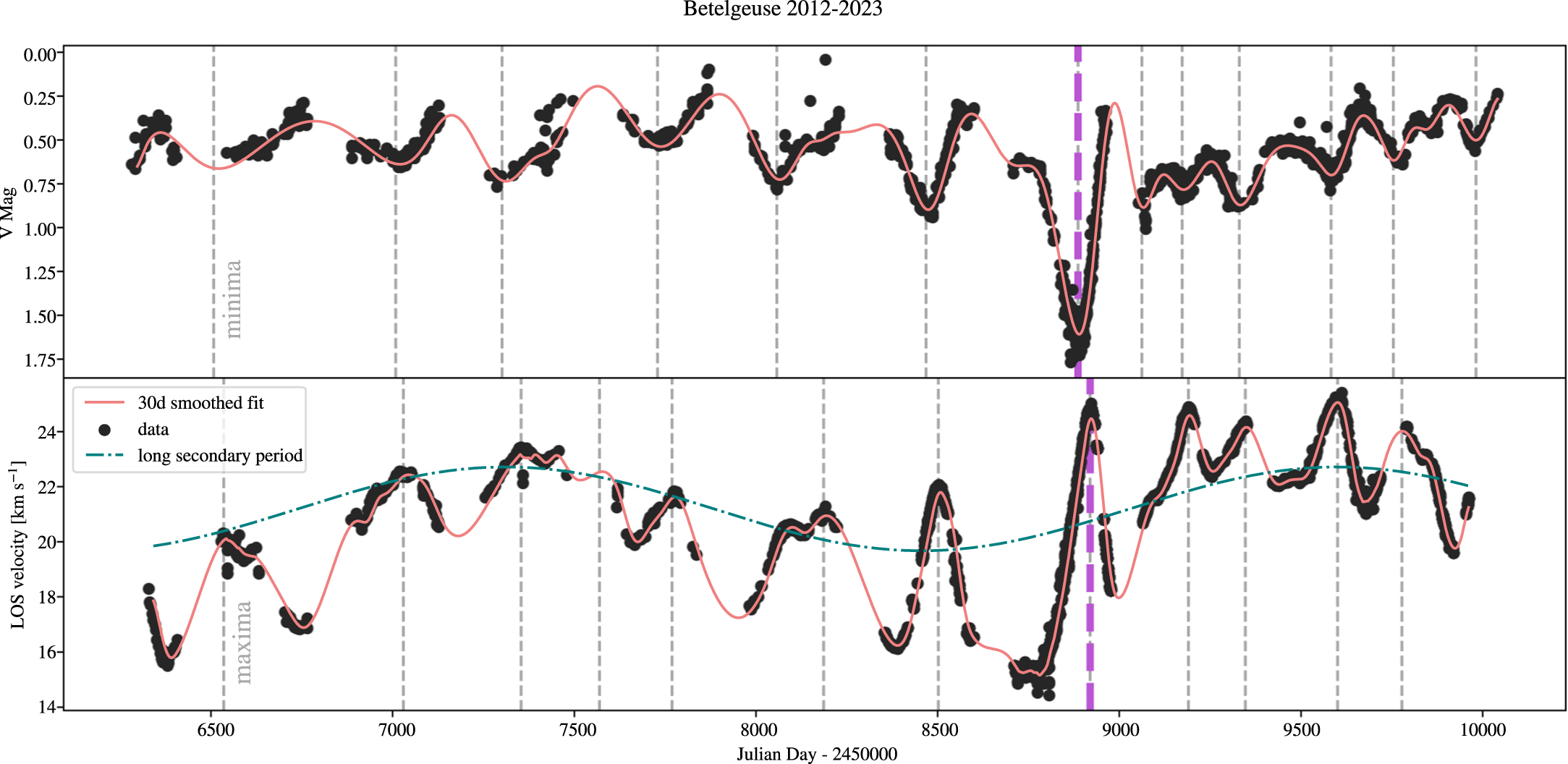}
	\caption{V-band photometry from the AAVSO (\textbf{upper panel}) and
	radial velocity (\textbf{lower panel}). From STELLA~\cite{2022csss.confE.185G}. The~red curve marks a
	smoothed spline fit to the observational data where the
	short period variation (300--400 days) is obvious prior to the Great Dimming.  
	Maxima and minima are identified by dashed vertical black lines. The~AAVSO data includes daytime observations
	from the ground acquired through new techniques which span the times when Betelgeuse is not available  at night. The~green dash-dot curve in the
	lower panel marks the long secondary period of about 2100~days. The~broken purple line marks the extremes in magnitude and velocity corresponding to the Great Dimming. The~average photospheric velocity is 20.6 km s$^{-1}$. Figure from~\cite{2023ApJ...956...27M}.\label{Fig:AAVSO_V}}
\end{figure}

Spectra in the optical region reveal the disk-averaged dynamics, and provide a probe of the
sub-surface plasma. Spectral lines arise from different levels throughout 
the photosphere and these can be used as depth-dependent velocity diagnostics. `Tomographic analysis' was used to detect a succession of two shocks in the photosphere  in February~2018 and January~2019 \cite{2021A&A...650L..17K}. Additionally, during~most of 2019,  the~average radial velocity of
the photosphere maintained a constant maximum outflowing value of about $-$6 km s$^{-1}$ as can be seen in Figure~\ref{Fig:AAVSO_V} \cite{2021csss.confE..41G}.  The~combination
of convection and the outward motion of the photosphere led to an ejection of plasma from the photosphere. The~lower chromosphere producing Ca II emission
also showed signatures of outflowing material at this time~\cite{2022ApJ...936...18D}.  Hydrodynamic simulations demonstrate~\cite{2023ApJ...956...27M} how a hot plume of gas moves
through the interior of a supergiant star, breaks the photosphere, and~extends over the stellar surface (Figure~\ref{Fig:Simulation}).  This leads to a Surface Mass Ejection (SME) and also 
breaks the phase coherence of the star's fundamental pulsation, leading to an overtone oscillation.  Such dramatic shortening of the 400-day pulsation is seen in the
optical magnitudes and the radial velocity (Figure~\ref{Fig:AAVSO_V}) after the Great~Dimming.

During 2019 and 2020, spatially resolved spectra of the chromosphere were obtained by HST/STIS (Figure~\ref{Fig:HST_2020})  with
an aperture of $25 \times 100$ mas, allowing about seven resolution elements across the stellar chromosphere~\cite{2020ApJ...899...68D}.
These spectra arise from higher layers in the star's chromosphere, ranging up to ~5 R$_{\star}$ \cite{1998AJ....116.2501U}.
While the spectra in the early months of 2019 did not appear unusual, during~2019 September--November, an~outflow
of plasma was detected in the Mg II chromospheric lines and C II indicated a density increase.  The~Mg II lines were also substantially enhanced by factors
of 3 or more in the Southern Hemisphere during this time.  A~delay of many months between a photospheric event and its appearance in the chromosphere is expected because of the
great size of the Betelgeuse atmosphere. However, during~the optical minimum, the~Mg~II flux returned to lower levels, reaching its lowest level 25 February 2020, perhaps obscured by dust in the~atmosphere.

\begin{figure}[H]

	\includegraphics[width=.95\textwidth]{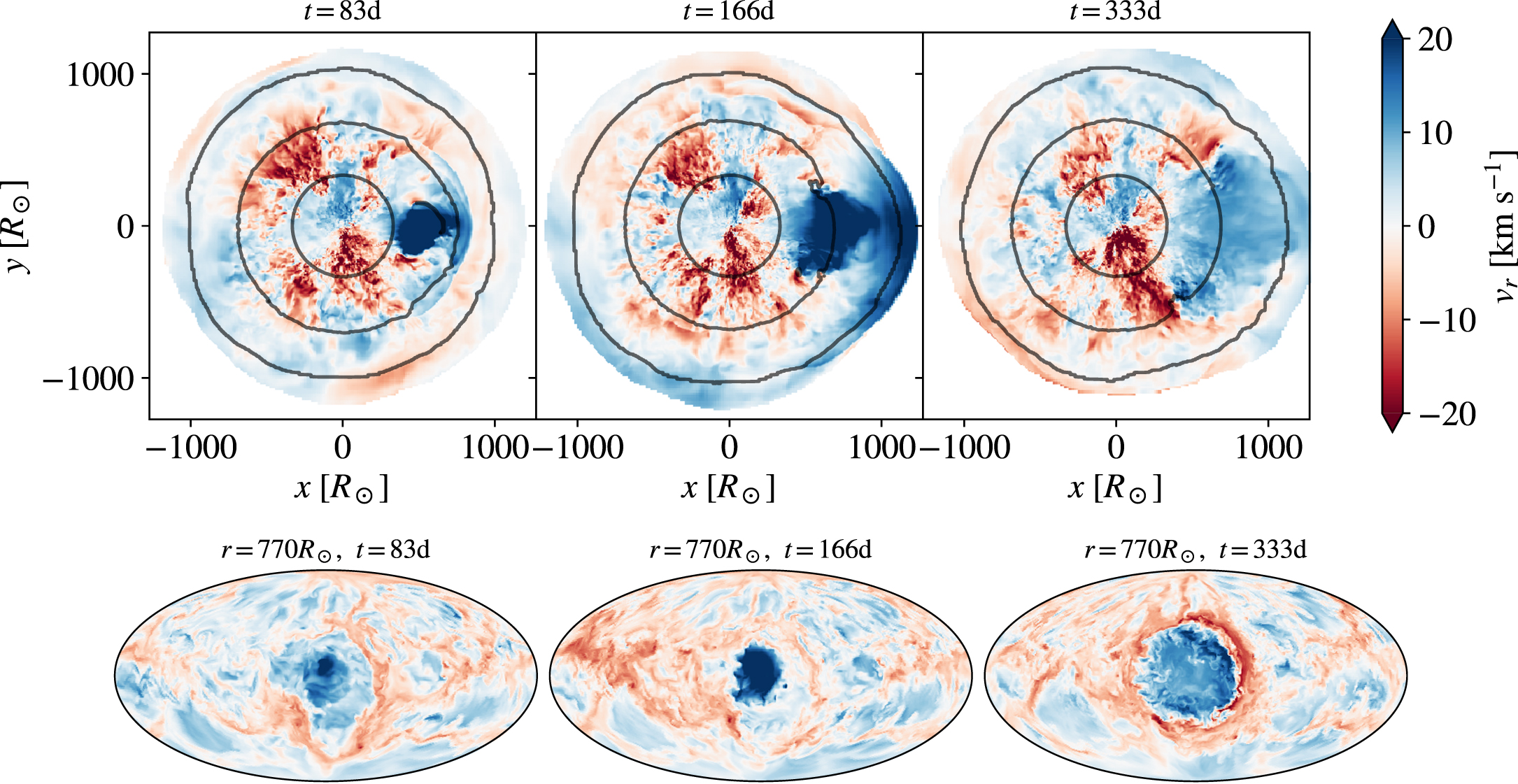}
	\caption{ Slices in radial velocity from the hydrodynamic calculations showing evolution of the hot plume of material as it approaches
		and spreads out over the stellar surface. Figure from~\cite{2023ApJ...956...27M}.
		\label{Fig:Simulation}}
\end{figure}
\unskip

\begin{figure}[H]
	
	\includegraphics[width=.48\textwidth]{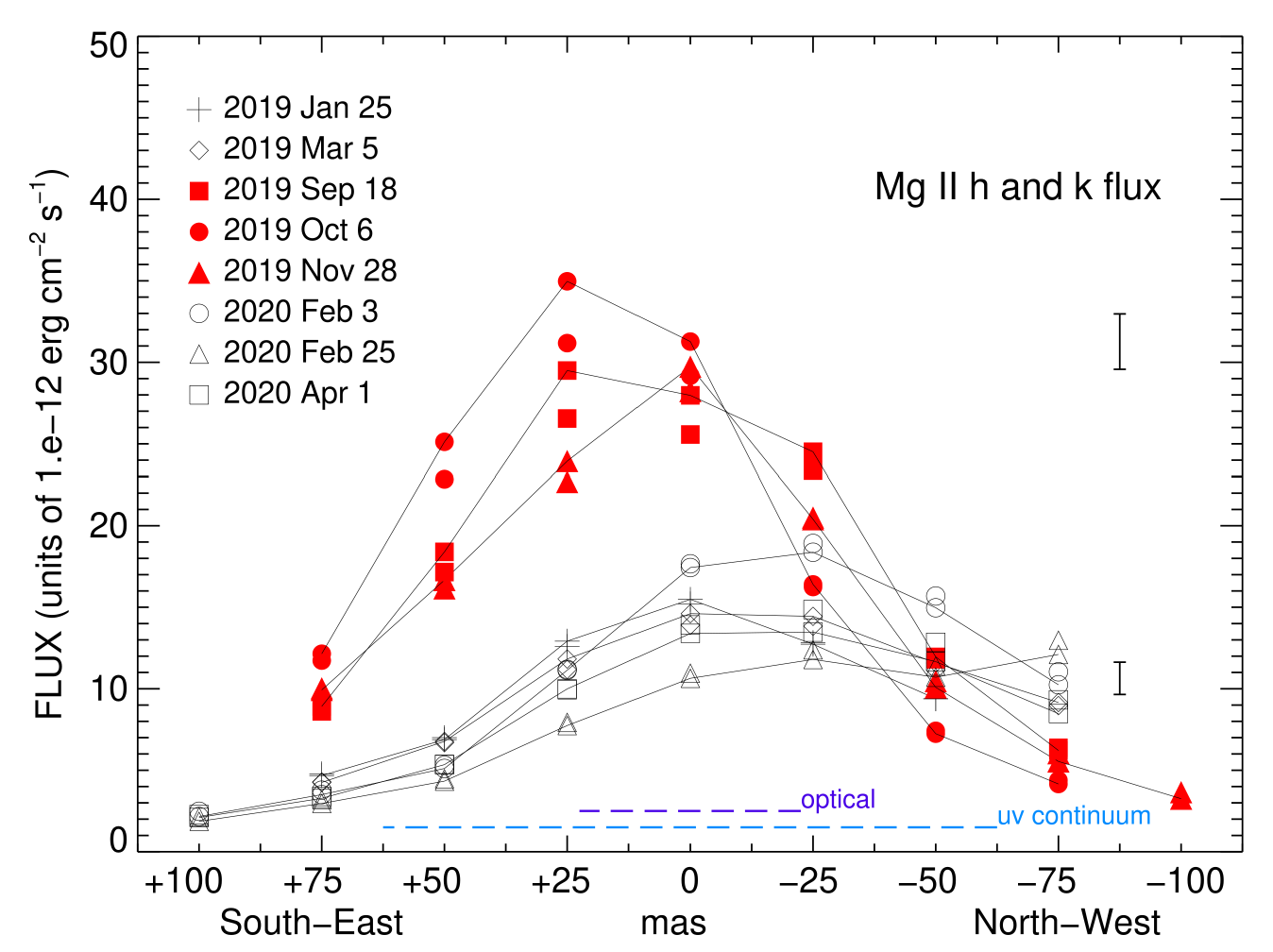}
	~
	\includegraphics[width=.48\textwidth]{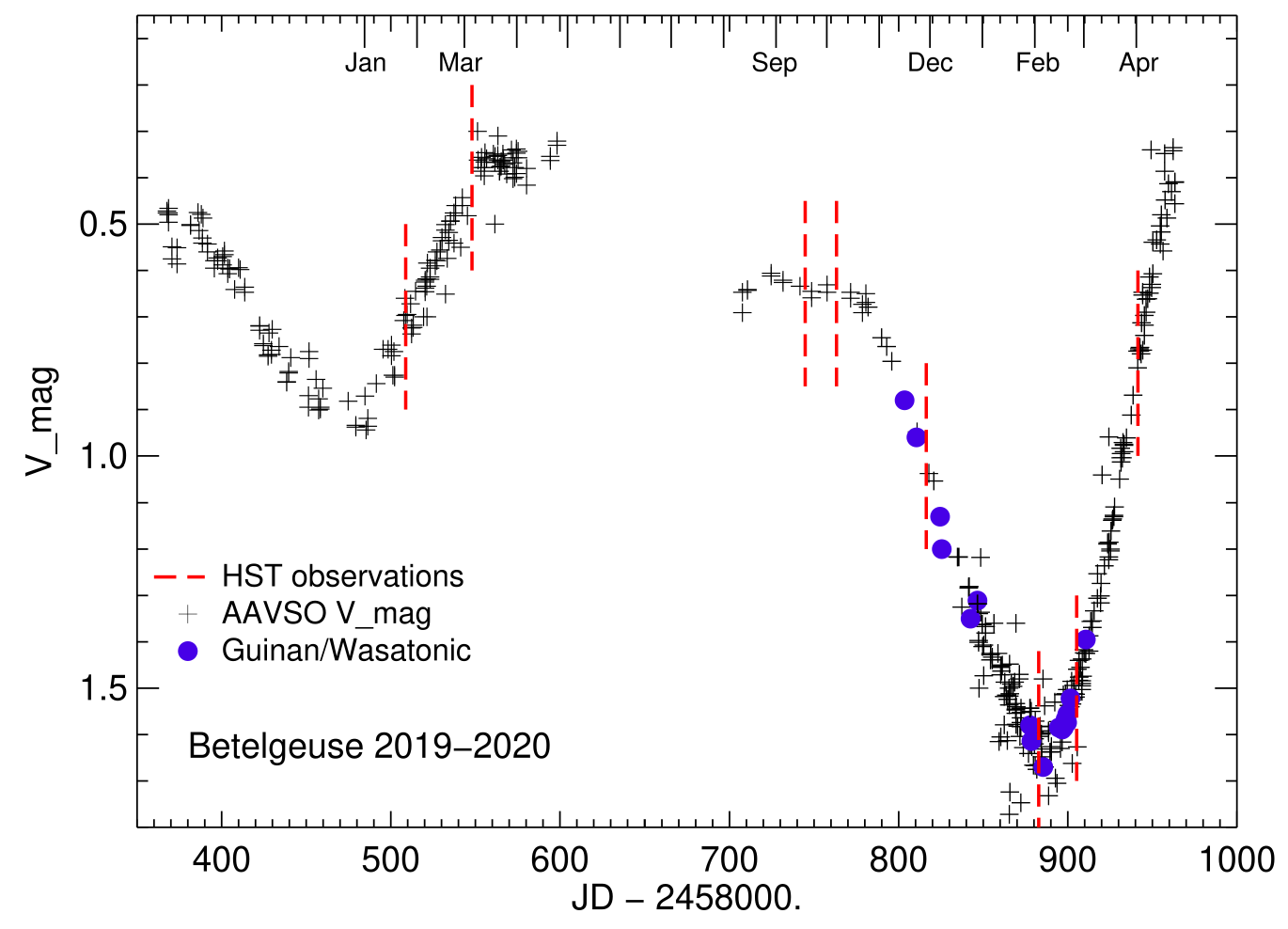}
	\caption{({\bf Left panel}) Mg II h and k fluxes (2803~\AA\  and 2795~\AA)  as measured from spatially resolved ultraviolet 
		STIS spectra from 2019 to 2020.  The~solid line connects the h-line of the doublet.  Note that the material ejected from 
		the surface in early 2019 did not reach the
		chromospheric levels until  about 6 months later and caused the strong enhancement in the southern hemisphere of the star.   The~chromospheric emission was the weakest on 25 February 2020,  
		decreasing at the star center by a factor of two near the optical
		minimum probably  caused by dust formation,  but~the Mg flux  recovered by 1 April 2020 to previous levels. ({\bf Right panel}) The V Magnitude with dates of STIS observations marked. Figure from~\cite{2020ApJ...899...68D}. \label{Fig:HST_2020}}
\end{figure}

The effects of this mass outflow appear to have cooled the photosphere and low chromosphere.  Increased  H$_2$O formation
was indicated at 6~$\upmu$m~\cite{2021MNRAS.502.4210T,2022NatAs...6..930T} and exceptionally low temperatures 2270~K (at 2.1 R$_\star$) and 2580~K (at 2.6 R$_\star$)  suggested by VLA measures at both millimeter and centimeter wavelengths~\cite{2022ApJ...934..131M}. Optical spectra and in particular, TiO bands
also supported a decrease in photospheric temperature to 3540 K--3645 K~\cite{2021A&A...650L..17K,2021NatCo..12.4719A,2019ATel13341....1G}. 

However, similar TiO band analyses showed that the decrease in temperature alone could not explain the Dimming~\cite{2020ApJ...891L..37L}. Dust was the designated next suspect. The~formation of dust in the line of sight, hence shadowing the stellar photosphere, has been suggested by classical polarimetric~\cite{2020RNAAS...4...39C} and speckle polarimetric interferometry~\cite{2020arXiv200505215S} observations.

Spatially resolved imaging of Betelgeuse has been obtained through the VLT/SPHERE adaptive optics instrument. The~images obtained a year before the Dimming (January 2019), and~throughout the event (fall in December 2019, minimum of brightness in January 2020, and~rise in March 2020, Figure~\ref{Fig:Dimming_SPHERE}) reveal a spectacular evolution of the visible ($\sim$655~nm) photosphere~\cite{2021Natur.594..365M}. Through \textsc{Phoenix} \cite{2007A&A...468..205L} and \textsc{Radmc3D} \cite{2012ascl.soft02015D} modeling, the~authors showed that both a cooling of the photosphere and dust formation in the line of sight were responsible for the~Dimming.  
	
	\vspace{-3pt}
\begin{figure}[H]
	
	\includegraphics[width=\textwidth]{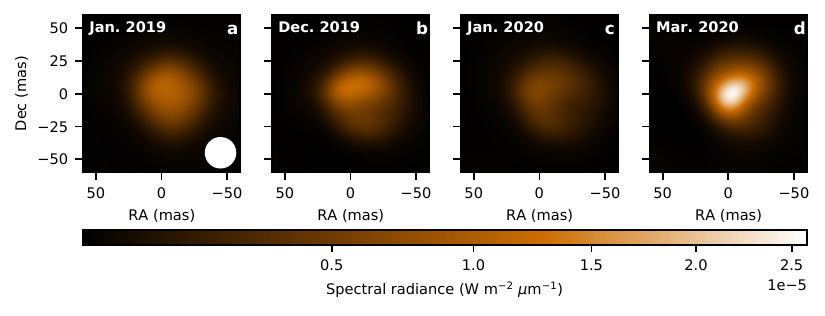}
	\caption{VLT/SPHERE adaptive optics imaging of the photosphere of Betelgeuse during the Great Dimming at 644.9~nm. North is up and East to the left. The~epoch of each observations is indicated in the top left corner of the images. The~white disk on the lower right corner of the first image shows the SPHERE point spread function. Figure from~\cite{2021Natur.594..365M}. \label{Fig:Dimming_SPHERE}}
\end{figure}

Chandra observations were attempted during the Great Dimming, in~order to possibly detect X-ray emissions from
the shocks that emerged from the photosphere or effects of a possible decrease in radius.  However, no detection was made in a 5.1 ks observation~\cite{2020ATel13501....1K}.

The Dimming of Betelgeuse has mobilized unusual resources to observe one of the most famous stars in our sky. Such an example is the Himawari-8 geostationary meteorological satellite. Such satellites offer the capability of delivering daily photometry of bright stars located near the celestial equator. This has been used to perform a multi-spectral (from 470~nm to 13.28~$\upmu$m) long-term (4.5~yr) monitoring of Betelgeuse (Figure~\ref{Fig:Taniguchi}, \cite{2022NatAs...6..930T}). The~simultaneous combination of wavelengths sensitive to temperature variations (visible domain sensitive to TiO bands) and dust presence (mid-infrared detecting the dust thermal emission) was the ideal tool for the authors to diagnose the Great Dimming. They confirm previous results~\cite{2021A&A...650L..17K,2021Natur.594..365M} that concluded a shared role of photospheric cooling and dust formation. Three-dimensional radiation-hydrodynamics simulations have finally supported the scenario involving cool gas and possibly dust~\cite{2024A&A...692A.223F}. However, this result is also a warning to observers: the observation wavelength is critical because it determines the atmospheric level being probed. The~ultraviolet region generally probes the chromosphere that can extend from the photosphere to several stellar radii; the optical region principally arises from the high photospheric levels, because~it is largely dominated by the TiO band opacities. The~near infrared probes a region closer to the photosphere (in terms of Rosseland opacity $\sim 1$) and the radio originates above the photosphere
and can extend to several stellar radii as well. The~Betelgeuse atmosphere that extends to several stellar radii is inhomogeneous and the selection of a wavelength to observe will probe the unique spatial scale, its velocity field and the evolution timescale of the plasma it represents~\cite{2024A&A...692A.223F}.


Another new resource for photometry resides in spacecraft designed for solar observations. NASA's Solar Terrestrial Relations Observatory (STEREO) consisted of two spacecraft launched orbiting the Sun in opposite directions---one ahead of the Earth and one behind. After~the Great Dimming, in~early 2020, STEREO-A was trailing the Earth in Earth's orbit by about 3 months.  It carried an Outer Heliospheric Imager as part of the SECCHI suite on STEREO, and~the spacecraft was rolled to repoint on the other side of the Sun and make photometric measurements that could be converted to V magnitudes~\cite{2015SoPh..290.2143T}.
The wide angle (70 degrees) of the field of view meant that other stars were available for calibration, and~its orbital position allowed STEREO-A to observe Betelgeuse when it could not be observed from the Earth~\cite{2020ATel13901....1D}.
Photometry prior to the Great Dimming was obtained from the Solar Mass Ejection Imager (SMEI) Space Experiment launched by the US Air Force and these measures were extracted to supplement the AAVSO measures a decade before the Great Dimming~\cite{2020ApJ...902...63J}.

A technique for obtaining photometry of Betelgeuse from the Earth during the daytime hours was
developed by an astronomer contributing to the AAVSO database (O. Nickel, 2020, contributing as NOT), and~these measures continue to contribute to give complete coverage of the star's behavior throughout the year. 
The great interest in Betelgeuse instigated a number of unique~observations.

\vspace{-4pt}
\begin{figure}[H]
	
	\includegraphics[width=\textwidth]{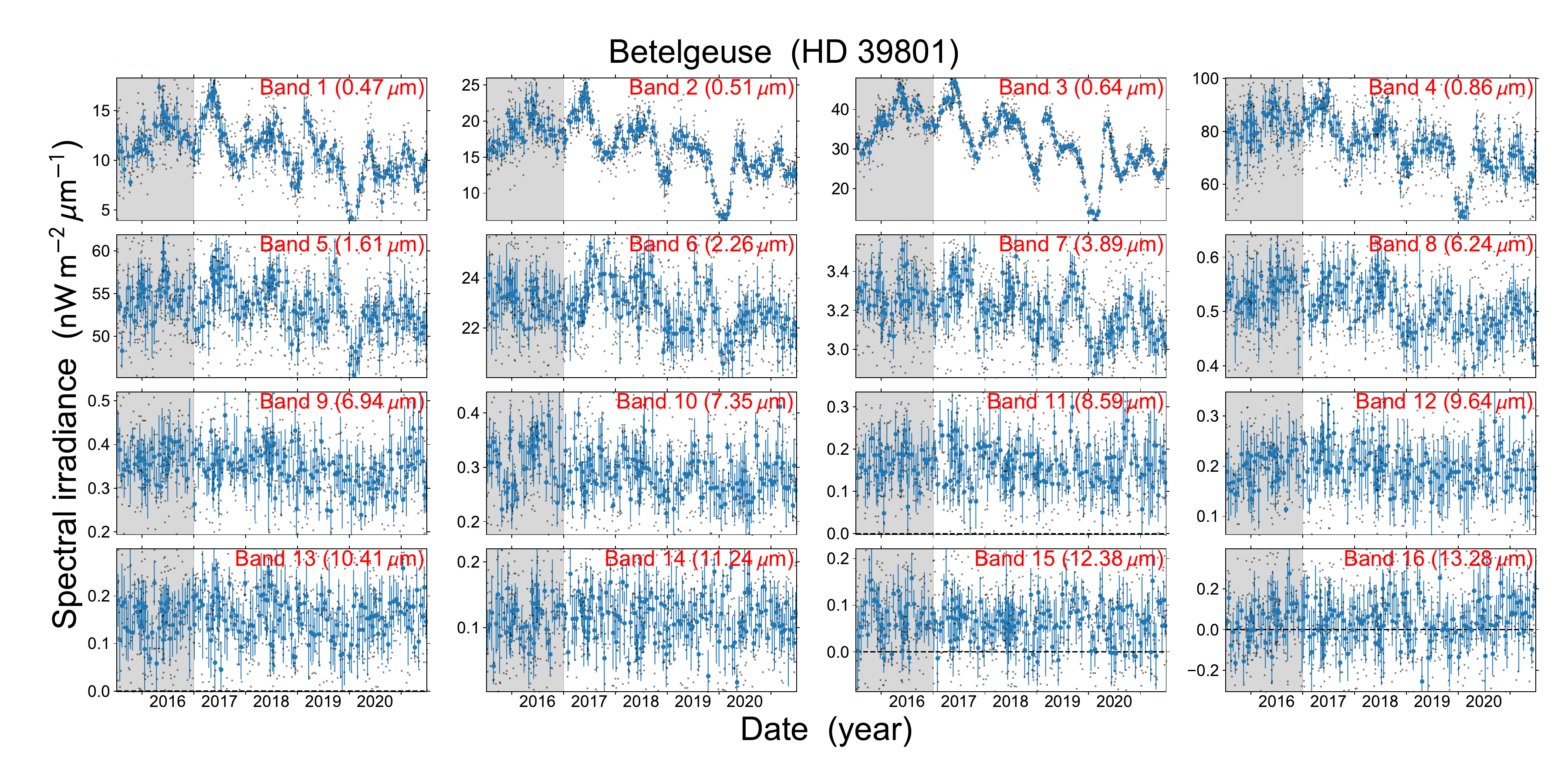}
	\caption{Sixteen-band monitoring of Betelgeuse over 4.5 years from 470~nm to 13.28~$\upmu$m using the weather satellite Himawari-8. Figure from~\cite{2022NatAs...6..930T}.\label{Fig:Taniguchi}}
\end{figure}
\unskip

\section{Long Secondary~Period\label{Sect:LSP}}

The Long Secondary Period (LSP) of Betelgeuse has been identified for some time with no conclusion as to its cause. Red giant stars  possess a LSP
and evidence has been put forth~\cite{2021ApJ...911L..22S} that many exhibit a secondary eclipse in the infrared which suggests  the presence of a dusty cloud that  may be associated
with a companion. Two recent publications~\cite{2024ApJ...977...35G,2025ApJ...978...50M} present extended analysis of Betelgeuse, with~different procedures, yet both concluded that the star
 hosts a companion. About ten of the  many different hypotheses for the LSP are scrutinized by~\cite{2024ApJ...977...35G}. They conclude the most attractive explanation is that of a 
 dusty body orbiting Betelgeuse. This hypothesis is consistent with the period of the optical light curve, the~amplitude of the stellar radial velocity, and~the delay between the radial velocity and the V-magnitude maxima. Another study~\cite{2025ApJ...978...50M} assembled a century of measurements of Betelgeuse: optical magnitude, radial velocity, and~astrometric studies. These quantities exhibit both periodic and aperiodic features. Analysis of the
 magnitudes and radial velocities incorporate periodograms and power spectra. The~LSP velocity variability is modeled in conjunction with a possible companion, and~demonstrates
 that the LSP is a stable signal in the radial velocity measures. Astrometry tends to favor the LSP and future measurements could give a stronger~result. 
 
 Characteristics of the companion object proposed in these studies are given in \mbox{Table~\ref{Tab:Companion}}~\cite{2024ApJ...977...35G,2025ApJ...978...50M}.  The~companion to 
 Betelgeuse  is  believed to be substantially fainter---perhaps a million times less luminous---than Betelgeuse, and it will be a challenge to detect directly.  The~separation of the faint companion is only about 50 mas from the center of Betelgeuse. However,
 requests for Director's Discretionary Time to search for a spectral signature, X-ray emission, or~an image of the companion have been granted for the 
 Hubble Space Telescope (M. Joyce), the~CHANDRA X-ray observatory (A. O'Grady), and~VLT/SPHERE (M. Montarg\`es). Even non-detections may be able to
 set interesting upper limits on the~companion.

\begin{table}[H] 
	\caption{Predicted values of the Betelgeuse binary~system. \label{Tab:Companion}}
	\newcolumntype{C}{>{\centering\arraybackslash}X}
	\begin{tabularx}{\textwidth}{CCC}
		\toprule
		\textbf{Property}	& \textbf{Value from~\cite{2024ApJ...977...35G}} & \textbf{Value from~\cite{2025ApJ...978...50M}} \\
		\midrule
		M$_a$ (M$_\odot$) & $18.0 \pm 1$    & $17.5 \pm 2$\\
		R$_a$ (R$_\odot$) & $764_{-62}^{+116}$& - \\
		M$_b$ (M$_\odot$) & $1.17 \pm 0.7$ & $0.60 \pm 0.14$  \\
		Period (days) & $2169 \pm 5.3$ & $2109 \pm 9$ \\
		T$_c$ (yr)  & 2023.45 $^{1}$ & $2023.12_{-0.35}^{+0.34}$ \\
		Separation (R$_\odot$) & $1850 \pm 70$ & $1818 \pm 6$ \\
		$\Omega$ ($^\circ$) & - & $60 \pm 6$ \\
		{\it i} ($^\circ$) & - & $98 \pm 5$  \\
		\bottomrule
	\end{tabularx}
	\noindent{\footnotesize{\textsuperscript{1} Phase 0, evaluated from the date of quadrature (phase 0.25).}}
\end{table}

These searches have taken place in November 2024 when one ephemeris~\cite{2024ApJ...977...35G} predicts quadrature, (phase 0.25) or the maximum elongation of the companion. It should be noted that another ephemeris~\cite{2025ApJ...978...50M} puts quadrature earlier by several months and suggests that the companion might be past
quadrature (phase 0.31) although there are uncertainties of $\pm 4$ months in specifying the phase.
The goal of the HST observation is to detect a spectral signature specific to the companion. Betelgeuse itself does not exhibit high temperature lines, arising from Si IV, C IV, etc. From~studies in the near and far ultraviolet spectral region, the~highest ions appear to be doubly ionized species: C II, Fe II, Cr II, Mg II etc,~\cite{1995AJ....109.2706B,2005ApJ...622..629D}. The~companion, with~a mass of $\sim 1$~M$_{\odot}$ or less could
be a K or M star or even later spectral type.  It might be accreting material from the stellar wind of the supergiant.
Spectra of such low mass stars frequently possess higher temperatures in their outer atmospheres whether accreting or not, and~so one might plausibly detect C IV or other species. The~separation of the companion, 1850~R$_{\odot}$ leads to an apparent separation from the center of Betelgeuse of 50~mas, and~clearly located within the extended stellar chromosphere. The~HST observations would likely point off the stellar limb to detect emission from the companion during~quadrature.

Time on ESO's VLT/SPHERE was also awarded for imaging in the optical at 644.9~nm. The~adaptive optics imaging (see Figure~\ref{Fig:Dimming_SPHERE}) clearly revealed the changes in the photosphere during the Dimming of Betelgeuse~\cite{2021Natur.594..365M}, and~might be expected to exhibit signs of a disturbance off the limb of the star. Using carefully selected filters, and~a pseudo-angular differential imaging technique, SPHERE is capable of reaching the $10^{-4}$ contrast required to detect the most massive predicted companion (A2V spectral type), even at less than 2~R$_\star$ from the photosphere. However, detecting the lightest companions (F8V) will not be feasible. More likely, these observations will determine an upper detection~limit.

\section{The Occultation by Solar System Asteroid 319 Leona in~2023}

In 2023, Betelgeuse was, again, at the center of attention. On 12 December, 
 an~occultation occurred when  (319) Leona, an~asteroid of our own Solar System, passed exactly in front of the star as seen from a tiny band across the Earth (Figure~\ref{Fig:Leona_occultation_map}).  When a spatially resolved asteroid passes in front of a distant star (considered as a point source), the~shadow of the asteroid crosses the Earth. The~silhouette of the asteroid can be reconstructed (see, e.g.,~\cite{2020A&A...639A.134O,2022A&A...663A.121V,2022A&A...664L..15M}). 
For~Betelgeuse, the~situation was not so simple: the characteristics that make it so precious for stellar physics made this occultation unique. The~angular size of Betelgeuse exceeded the angular size of the asteroid. In~other words, instead of an occultation, the~observers witnessed the transit of Leona in front of Betelgeuse. A~team (in French, \url{https://gemini.obspm.fr/20230715-betelgeuse/}, accessed on 21 April 2025})
 decided to coordinate the observational effort in an attempt to obtain an image of the surface of Betelgeuse from this event. The~full results of this campaign have yet to be published. An~experiment on a single location (a single chord) using a Single-Photon Avalanche Diode detected a 77.78\% occultation of the star and a visible photospheric diameter of Betelgeuse of 57.26~mas~\cite{2024arXiv240614704P}.

\begin{figure}[H]
	
	\includegraphics[width=\textwidth]{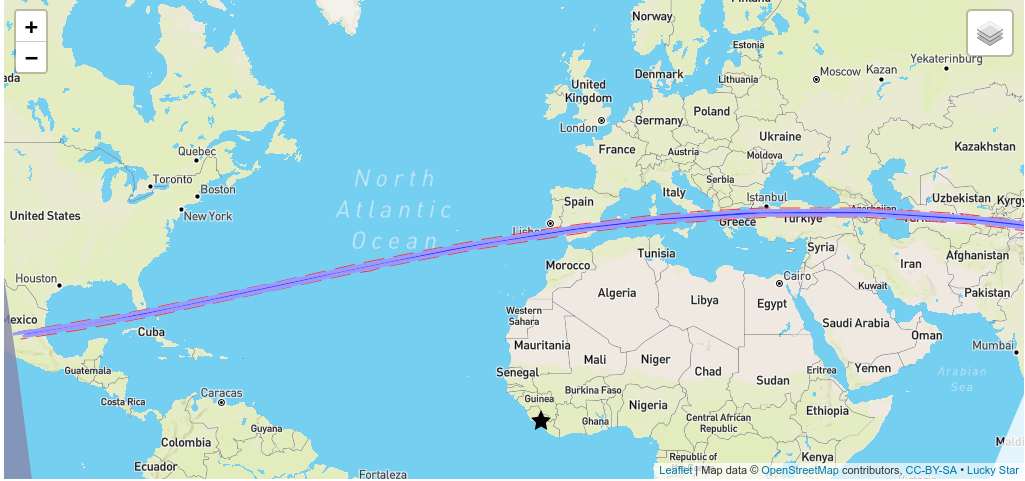}
	\caption{Map of the trace on the ground from the occultation of Betelgeuse by the solar system asteroid (319) Leona on 12 December 2023 (\url{https://lesia.obspm.fr/lucky-star/occ.php?p=131608}, accessed on 21 April 2025). \label{Fig:Leona_occultation_map}}
\end{figure}

\vspace{-22pt}
\section{Implications for Other Supergiant~Stars}

The size and brightness of Betelgeuse offer a unique view into its atmospheric conditions and variability. These insights 
complement the study of other supergiants where spatial resolution is not possible.  Foremost among these stars
is VY CMa which, although~dust obscures the star, gives evidence of episodic outflow in the form of infrared-bright knots and clumps found in the region surrounding the star~\cite{2019ApJ...874L..26H,2022AJ....163..103H}. The~location of these knots does not show any preferred direction
suggesting that the driving mechanism may be independent of a stellar characteristic such as stellar rotation and a magnetic field. Similar to Betelgeuse and the Great Dimming event, the~ejection times estimated for knots surrounding VY CMa appear to correspond with deep minima and extended times of variability~\cite{2021AJ....161...98H}. 
As noted previously, a~similar evaluation for Betelgeuse  did not reveal such correlations~\cite{2022AJ....163..103H}. 

The Great Dimming in 2019--2020 triggered much interest in the evolved stars community. It was not long before a similar event was found in the archival HST observations of the galaxy M51~\cite{2022ApJ...930...81J}. The~star, designated M51-DS1,  has colors consistent with a RSG of initial mass $\approx 19$--$22$~M$_\odot$. 
It experienced a 2~mag dimming between late 2017 and mid-2019, observed in the F814W filter of the Advanced Camera for Survey (ACS)/Wide Field Camera 3 (AFC3) of the HST. This is significantly longer than the few months of Betelgeuse's, and~has been interpreted as a more important enhanced mass-loss~episode.  

Another exceptional event has been detected on the yellow hypergiant RW~Cep in late 2022. The~changing photosphere has been observed through optical long baseline interferometry with the CHARA array using the MIRC-X instrument~\cite{2023AJ....166...78A}. The~multi-spectral analysis concluded that the star had experienced an obscuration by newly formed dust. This has been confirmed by a second study monitoring the star during the re-brightening~\cite{2024ApJ...973L...5A}. This dimming presents differences with the Great Dimming of Betelgeuse. 
From~RW~Cep's light curve (\cite{2023AJ....166...78A}, Figure~1),
 it appears that the star has been slowly dimming over its previous pulsation cycle (2~years before the actual event). In~contrast, the~Dimming of Betelgeuse has been consistent with a single pulsation~cycle.

In the Large Magellanic Cloud, a~very luminous red supergiant [W60]B90 exhibits a bow shock~\cite{2024A&A...690A..99M} reminiscent of that
observed to be associated with Betelgeuse and attributed in part to a clumpy mass-loss process~\cite{2012A&A...548A.113D}. The light curves of [W60]B90  indicate three dimming events over almost 24 years, accompanied by a decrease in atmospheric temperature indicated by changes in color and spectra~\cite{2024A&A...690A..99M}. 

All of these measurements contribute to  indicate that an episodic mass loss process  may be a common phenomenon among red supergiants.
Interestingly, instead of having been a singular event, the~Great Dimming of Betelgeuse, a~surface mass ejection, has triggered the hunt for similar occurrences on other cool evolved stars. Betelgeuse does not appear isolated, and~such enhanced mass-loss episodes could be more common than initially thought. Perhaps we only needed Betelgeuse to blink at us to realize~this.

\section{Conclusions}

After millennia of naked eye observations, and~a few centuries of telescope scrutinizing, Betelgeuse is still a keystone of modern astronomy. The~Great Dimming in 2019--2020 has shown us that even without going supernova, a~tiny blink of its luminosity can trigger a worldwide observing campaign, with~much interest from the public in addition to astronomers. The~observation of this nearby prototypical star paves the way for a better understanding of the mass loss of~RSGs.

In the months following this review, the~outcome of the observing campaign searching for Betelgeuse B may become known. No doubt this quest, successful or only posing upper detection limits, will have a strong impact on the evolutionary status of Betelgeuse. As~usual, anything that is better constrained on Betelgeuse will have wider implications for the RSG population. Hopefully, the~coming years will see additional insights emerging from our attempts at constraining the distance and rotation velocity of this prototypical~star. 

Observing facilities under construction are usually dedicated to pushing the limit of sensitivity. We should not forget that larger telescopes are also instruments with  higher angular resolution. For~the past century, Betelgeuse has been at the forefront of stellar physics in terms of spatially resolved details being detected and analyzed. Despite its constraining brightness, it should remain a prime target for observers. After~all, it would be best to learn everything we can before it becomes the second-brightest star in Earth skies, when it will explode as a~supernova.



\vspace{6pt} 





\vspace{6pt}
\authorcontributions{Conceptualization, A.K.D. and M.M.; validation, A.K.D. and M.M.; investigation, A.K.D and M.M;  writing---original draft preparation, A.K.D. and M.M.; writing---review and editing, A.K.D. and M.M.; All authors have read and agreed to the published version of the manuscript.}

\funding{This research received no external~funding}




\conflictsofinterest{The authors declare no conflicts of~interest.} 



\appendixtitles{no} 
%
%

\begin{adjustwidth}{-\extralength}{0cm}

\reftitle{References}

\PublishersNote{}
\end{adjustwidth}

\begin{thebibliography}{999}

\bibitem[{Rappengl{\"u}ck}(2003)]{2003UppOR..59...51R}
{Rappengl{\"u}ck}, M.
\newblock {The anthropoid in the sky: Does a 32,000-year old ivory plate show
  the constellation Orion combined with a pregnancy calendar?}
\newblock {\em Uppsala Astronomical Observatory Reports} {\bf 2003}, {\em
  59},~51.

\bibitem[{Neuh{\"a}user} et~al.(2022){Neuh{\"a}user}, {Torres}, {Mugrauer},
  {Neuh{\"a}user}, {Chapman}, {Luge}, and {Cosci}]{2022MNRAS.516..693N}
{Neuh{\"a}user}, R.; {Torres}, G.; {Mugrauer}, M.; {Neuh{\"a}user}, D.L.;
  {Chapman}, J.; {Luge}, D.; {Cosci}, M.
\newblock {Colour evolution of Betelgeuse and Antares over two millennia,
  derived from historical records, as a new constraint on mass and age}.
\newblock {\em \mnras} {\bf 2022}, {\em 516},~693--719,
  \href{http://arxiv.org/abs/2207.04702}{{\normalfont
  [arXiv:astro-ph.SR/2207.04702]}}.
\newblock {\url{https://doi.org/10.1093/mnras/stac1969}}.

\bibitem[{Michelson} and {Pease}(1921)]{1921ApJ....53..249M}
{Michelson}, A.A.; {Pease}, F.G.
\newblock {Measurement of the diameter of alpha Orionis with the
  interferometer.}
\newblock {\em \apj} {\bf 1921}, {\em 53},~249--259.
\newblock {\url{https://doi.org/10.1086/142603}}.

\bibitem[NYt(1920)]{NYtimes1920}
GIANT STAR EQUAL TO 27,000,000 SUNS LIKE OURS.
\newblock {\em The New York Times} {\bf 1920}.

\bibitem[{Kervella} et~al.(2013){Kervella}, {Le Bertre}, and
  {Perrin}]{2013EAS....60.....K}
{Kervella}, P.; {Le Bertre}, T.; {Perrin}, G., Eds.
\newblock {\em {Betelgeuse Workshop 2012}}, Vol.~60, {\em EAS Publications
  Series},  2013.

\bibitem[{Keenan} and {McNeil}(1989)]{1989ApJS...71..245K}
{Keenan}, P.C.; {McNeil}, R.C.
\newblock {The Perkins Catalog of Revised MK Types for the Cooler Stars}.
\newblock {\em \apjs} {\bf 1989}, {\em 71},~245.
\newblock {\url{https://doi.org/10.1086/191373}}.

\bibitem[{Hoffleit} and {Warren}(1987)]{1987ADCBu...1..285H}
{Hoffleit}, D.; {Warren}, Jr., W.H.
\newblock {The Bright Star Catalogue, 5$^{th}$ revised edition.}
\newblock {\em Astronomical Data Center Bulletin} {\bf 1987}, {\em
  1},~285--294.

\bibitem[{Gilliland} and {Dupree}(1996)]{1996ApJ...463L..29G}
{Gilliland}, R.L.; {Dupree}, A.K.
\newblock {First Image of the Surface of a Star with the Hubble Space
  Telescope}.
\newblock {\em \apjl} {\bf 1996}, {\em 463},~L29.
\newblock {\url{https://doi.org/10.1086/310043}}.

\bibitem[{Ohnaka} et~al.(2009){Ohnaka}, {Hofmann}, {Benisty}, {Chelli},
  {Driebe}, {Millour}, {Petrov}, {Schertl}, {Stee}, {Vakili}, and
  {Weigelt}]{2009A&A...503..183O}
{Ohnaka}, K.; {Hofmann}, K.H.; {Benisty}, M.; {Chelli}, A.; {Driebe}, T.;
  {Millour}, F.; {Petrov}, R.; {Schertl}, D.; {Stee}, P.; {Vakili}, F.;  et~al.
\newblock {Spatially resolving the inhomogeneous structure of the dynamical
  atmosphere of Betelgeuse with VLTI/AMBER}.
\newblock {\em \aap} {\bf 2009}, {\em 503},~183--195,
  \href{http://arxiv.org/abs/0906.4792}{{\normalfont
  [arXiv:astro-ph.SR/0906.4792]}}.
\newblock {\url{https://doi.org/10.1051/0004-6361/200912247}}.

\bibitem[{Haubois} et~al.(2009){Haubois}, {Perrin}, {Lacour}, {Verhoelst},
  {Meimon}, {Mugnier}, {Thi{\'e}baut}, {Berger}, {Ridgway}, {Monnier},
  {Millan-Gabet}, and {Traub}]{2009A&A...508..923H}
{Haubois}, X.; {Perrin}, G.; {Lacour}, S.; {Verhoelst}, T.; {Meimon}, S.;
  {Mugnier}, L.; {Thi{\'e}baut}, E.; {Berger}, J.P.; {Ridgway}, S.T.;
  {Monnier}, J.D.;  et~al.
\newblock {Imaging the spotty surface of <ASTROBJ>Betelgeuse</ASTROBJ> in the H
  band}.
\newblock {\em \aap} {\bf 2009}, {\em 508},~923--932,
  \href{http://arxiv.org/abs/0910.4167}{{\normalfont
  [arXiv:astro-ph.SR/0910.4167]}}.
\newblock {\url{https://doi.org/10.1051/0004-6361/200912927}}.

\bibitem[{Ohnaka} et~al.(2011){Ohnaka}, {Weigelt}, {Millour}, {Hofmann},
  {Driebe}, {Schertl}, {Chelli}, {Massi}, {Petrov}, and
  {Stee}]{2011A&A...529A.163O}
{Ohnaka}, K.; {Weigelt}, G.; {Millour}, F.; {Hofmann}, K.H.; {Driebe}, T.;
  {Schertl}, D.; {Chelli}, A.; {Massi}, F.; {Petrov}, R.; {Stee}, P.
\newblock {Imaging the dynamical atmosphere of the red supergiant Betelgeuse in
  the CO first overtone lines with VLTI/AMBER}.
\newblock {\em \aap} {\bf 2011}, {\em 529},~A163,
  \href{http://arxiv.org/abs/1104.0958}{{\normalfont
  [arXiv:astro-ph.SR/1104.0958]}}.
\newblock {\url{https://doi.org/10.1051/0004-6361/201016279}}.

\bibitem[{Dupree} and {Stefanik}(2013)]{2013EAS....60...77D}
{Dupree}, A.K.; {Stefanik}, R.P.
\newblock {Direct ultraviolet imaging and spectroscopy of betelgeuse}.
\newblock In Proceedings of the EAS Publications Series; {Kervella}, P.; {Le
  Bertre}, T.; {Perrin}, G., Eds. EDP,  2013, Vol.~60, {\em EAS Publications
  Series}, pp. 77--84,  \href{http://arxiv.org/abs/1304.2780}{{\normalfont
  [arXiv:astro-ph.SR/1304.2780]}}.
\newblock {\url{https://doi.org/10.1051/eas/1360008}}.

\bibitem[{Montarg{\`e}s} et~al.(2016){Montarg{\`e}s}, {Kervella}, {Perrin},
  {Chiavassa}, {Le Bouquin}, {Auri{\`e}re}, {L{\'o}pez Ariste}, {Mathias},
  {Ridgway}, {Lacour}, {Haubois}, and {Berger}]{2016A&A...588A.130M}
{Montarg{\`e}s}, M.; {Kervella}, P.; {Perrin}, G.; {Chiavassa}, A.; {Le
  Bouquin}, J.B.; {Auri{\`e}re}, M.; {L{\'o}pez Ariste}, A.; {Mathias}, P.;
  {Ridgway}, S.T.; {Lacour}, S.;  et~al.
\newblock {The close circumstellar environment of Betelgeuse. IV. VLTI/PIONIER
  interferometric monitoring of the photosphere}.
\newblock {\em \aap} {\bf 2016}, {\em 588},~A130,
  \href{http://arxiv.org/abs/1602.05108}{{\normalfont
  [arXiv:astro-ph.SR/1602.05108]}}.
\newblock {\url{https://doi.org/10.1051/0004-6361/201527028}}.

\bibitem[{O'Gorman} et~al.(2017){O'Gorman}, {Kervella}, {Harper}, {Richards},
  {Decin}, {Montarg{\`e}s}, and {McDonald}]{2017A&A...602L..10O}
{O'Gorman}, E.; {Kervella}, P.; {Harper}, G.M.; {Richards}, A.M.S.; {Decin},
  L.; {Montarg{\`e}s}, M.; {McDonald}, I.
\newblock {The inhomogeneous submillimeter atmosphere of Betelgeuse}.
\newblock {\em \aap} {\bf 2017}, {\em 602},~L10,
  \href{http://arxiv.org/abs/1706.06021}{{\normalfont
  [arXiv:astro-ph.SR/1706.06021]}}.
\newblock {\url{https://doi.org/10.1051/0004-6361/201731171}}.

\bibitem[{L{\'o}pez Ariste} et~al.(2022){L{\'o}pez Ariste}, {Georgiev},
  {Mathias}, {L{\`e}bre}, {Wavasseur}, {Josselin}, {Konstantinova-Antova}, and
  {Roudier}]{2022A&A...661A..91L}
{L{\'o}pez Ariste}, A.; {Georgiev}, S.; {Mathias}, P.; {L{\`e}bre}, A.;
  {Wavasseur}, M.; {Josselin}, E.; {Konstantinova-Antova}, R.; {Roudier}, T.
\newblock {Three-dimensional imaging of convective cells in the photosphere of
  Betelgeuse}.
\newblock {\em \aap} {\bf 2022}, {\em 661},~A91,
  \href{http://arxiv.org/abs/2202.12011}{{\normalfont
  [arXiv:astro-ph.SR/2202.12011]}}.
\newblock {\url{https://doi.org/10.1051/0004-6361/202142271}}.

\bibitem[{Pilate} et~al.(2024){Pilate}, {Ariste}, {Lavail}, and
  {Mathias}]{2024A&A...691A.297P}
{Pilate}, Q.; {Ariste}, A.L.; {Lavail}, A.; {Mathias}, P.
\newblock {The variability of Betelgeuse explained by surface convection}.
\newblock {\em \aap} {\bf 2024}, {\em 691},~A297,
  \href{http://arxiv.org/abs/2410.08819}{{\normalfont
  [arXiv:astro-ph.SR/2410.08819]}}.
\newblock {\url{https://doi.org/10.1051/0004-6361/202450987}}.

\bibitem[{Chiavassa} et~al.(2010){Chiavassa}, {Haubois}, {Young}, {Plez},
  {Josselin}, {Perrin}, and {Freytag}]{2010A&A...515A..12C}
{Chiavassa}, A.; {Haubois}, X.; {Young}, J.S.; {Plez}, B.; {Josselin}, E.;
  {Perrin}, G.; {Freytag}, B.
\newblock {Radiative hydrodynamics simulations of red supergiant stars. II.
  Simulations of convection on Betelgeuse match interferometric observations}.
\newblock {\em \aap} {\bf 2010}, {\em 515},~A12,
  \href{http://arxiv.org/abs/1003.1407}{{\normalfont
  [arXiv:astro-ph.SR/1003.1407]}}.
\newblock {\url{https://doi.org/10.1051/0004-6361/200913907}}.

\bibitem[{Montarg{\`e}s} et~al.(2021){Montarg{\`e}s}, {Cannon}, {Lagadec}, {de
  Koter}, {Kervella}, {Sanchez-Bermudez}, {Paladini}, {Cantalloube}, {Decin},
  {Scicluna}, {Kravchenko}, {Dupree}, {Ridgway}, {Wittkowski}, {Anugu},
  {Norris}, {Rau}, {Perrin}, {Chiavassa}, {Kraus}, {Monnier}, {Millour}, {Le
  Bouquin}, {Haubois}, {Lopez}, {Stee}, and {Danchi}]{2021Natur.594..365M}
{Montarg{\`e}s}, M.; {Cannon}, E.; {Lagadec}, E.; {de Koter}, A.; {Kervella},
  P.; {Sanchez-Bermudez}, J.; {Paladini}, C.; {Cantalloube}, F.; {Decin}, L.;
  {Scicluna}, P.;  et~al.
\newblock {A dusty veil shading Betelgeuse during its Great Dimming}.
\newblock {\em \nat} {\bf 2021}, {\em 594},~365--368.
\newblock {\url{https://doi.org/10.1038/s41586-021-03546-8}}.

\bibitem[{Kervella} et~al.(2009){Kervella}, {Verhoelst}, {Ridgway}, {Perrin},
  {Lacour}, {Cami}, and {Haubois}]{2009A&A...504..115K}
{Kervella}, P.; {Verhoelst}, T.; {Ridgway}, S.T.; {Perrin}, G.; {Lacour}, S.;
  {Cami}, J.; {Haubois}, X.
\newblock {The close circumstellar environment of Betelgeuse. Adaptive optics
  spectro-imaging in the near-IR with VLT/NACO}.
\newblock {\em \aap} {\bf 2009}, {\em 504},~115--125,
  \href{http://arxiv.org/abs/0907.1843}{{\normalfont
  [arXiv:astro-ph.SR/0907.1843]}}.
\newblock {\url{https://doi.org/10.1051/0004-6361/200912521}}.

\bibitem[{Decin} et~al.(2012){Decin}, {Cox}, {Royer}, {Van Marle},
  {Vandenbussche}, {Ladjal}, {Kerschbaum}, {Ottensamer}, {Barlow}, {Blommaert},
  {Gomez}, {Groenewegen}, {Lim}, {Swinyard}, {Waelkens}, and
  {Tielens}]{2012A&A...548A.113D}
{Decin}, L.; {Cox}, N.L.J.; {Royer}, P.; {Van Marle}, A.J.; {Vandenbussche},
  B.; {Ladjal}, D.; {Kerschbaum}, F.; {Ottensamer}, R.; {Barlow}, M.J.;
  {Blommaert}, J.A.D.L.;  et~al.
\newblock {The enigmatic nature of the circumstellar envelope and bow shock
  surrounding Betelgeuse as revealed by Herschel. I. Evidence of clumps,
  multiple arcs, and a linear bar-like structure}.
\newblock {\em \aap} {\bf 2012}, {\em 548},~A113,
  \href{http://arxiv.org/abs/1212.4870}{{\normalfont
  [arXiv:astro-ph.SR/1212.4870]}}.
\newblock {\url{https://doi.org/10.1051/0004-6361/201219792}}.

\bibitem[{Wheeler} and {Chatzopoulos}(2023)]{2023A&G....64.3.11W}
{Wheeler}, J.C.; {Chatzopoulos}, E.
\newblock {Betelgeuse: a review}.
\newblock {\em Astronomy and Geophysics} {\bf 2023}, {\em 64},~3.11--3.27,
  \href{http://arxiv.org/abs/2306.09449}{{\normalfont
  [arXiv:astro-ph.SR/2306.09449]}}.
\newblock {\url{https://doi.org/10.1093/astrogeo/atad020}}.

\bibitem[{Uitenbroek} et~al.(1998){Uitenbroek}, {Dupree}, and
  {Gilliland}]{1998AJ....116.2501U}
{Uitenbroek}, H.; {Dupree}, A.K.; {Gilliland}, R.L.
\newblock {Spatially Resolved Hubble Space Telescope Spectra of the
  Chromosphere of alpha Orionis}.
\newblock {\em \aj} {\bf 1998}, {\em 116},~2501--2512.
\newblock {\url{https://doi.org/10.1086/300596}}.

\bibitem[{Dupree} et~al.(2020){Dupree}, {Strassmeier}, {Matthews},
  {Uitenbroek}, {Calderwood}, {Granzer}, {Guinan}, {Leike}, {Montarg{\`e}s},
  {Richards}, {Wasatonic}, and {Weber}]{2020ApJ...899...68D}
{Dupree}, A.K.; {Strassmeier}, K.G.; {Matthews}, L.D.; {Uitenbroek}, H.;
  {Calderwood}, T.; {Granzer}, T.; {Guinan}, E.F.; {Leike}, R.;
  {Montarg{\`e}s}, M.; {Richards}, A.M.S.;  et~al.
\newblock {Spatially Resolved Ultraviolet Spectroscopy of the Great Dimming of
  Betelgeuse}.
\newblock {\em \apj} {\bf 2020}, {\em 899},~68,
  \href{http://arxiv.org/abs/2008.04945}{{\normalfont
  [arXiv:astro-ph.SR/2008.04945]}}.
\newblock {\url{https://doi.org/10.3847/1538-4357/aba516}}.

\bibitem[{Wilson} et~al.(1997){Wilson}, {Dhillon}, and
  {Haniff}]{1997MNRAS.291..819W}
{Wilson}, R.W.; {Dhillon}, V.S.; {Haniff}, C.A.
\newblock {The changing face of Betelgeuse}.
\newblock {\em \mnras} {\bf 1997}, {\em 291},~819--826.
\newblock {\url{https://doi.org/10.1093/mnras/291.4.819}}.

\bibitem[{Schwarzschild}(1975)]{1975ApJ...195..137S}
{Schwarzschild}, M.
\newblock {On the scale of photospheric convection in red giants and
  supergiants}.
\newblock {\em \apj} {\bf 1975}, {\em 195},~137--144.
\newblock {\url{https://doi.org/10.1086/153313}}.

\bibitem[{Kravchenko} et~al.(2021){Kravchenko}, {Jorissen}, {Van Eck}, {Merle},
  {Chiavassa}, {Paladini}, {Freytag}, {Plez}, {Montarg{\`e}s}, and {Van
  Winckel}]{2021A&A...650L..17K}
{Kravchenko}, K.; {Jorissen}, A.; {Van Eck}, S.; {Merle}, T.; {Chiavassa}, A.;
  {Paladini}, C.; {Freytag}, B.; {Plez}, B.; {Montarg{\`e}s}, M.; {Van
  Winckel}, H.
\newblock {Atmosphere of Betelgeuse before and during the Great Dimming event
  revealed by tomography}.
\newblock {\em \aap} {\bf 2021}, {\em 650},~L17,
  \href{http://arxiv.org/abs/2104.08105}{{\normalfont
  [arXiv:astro-ph.SR/2104.08105]}}.
\newblock {\url{https://doi.org/10.1051/0004-6361/202039801}}.

\bibitem[{Jadlovsk{\'y}} et~al.(2024){Jadlovsk{\'y}}, {Granzer}, {Weber},
  {Kravchenko}, {Krti{\v{c}}ka}, {Dupree}, {Chiavassa}, {Strassmeier}, and
  {Poppenh{\"a}ger}]{2024A&A...685A.124J}
{Jadlovsk{\'y}}, D.; {Granzer}, T.; {Weber}, M.; {Kravchenko}, K.;
  {Krti{\v{c}}ka}, J.; {Dupree}, A.K.; {Chiavassa}, A.; {Strassmeier}, K.G.;
  {Poppenh{\"a}ger}, K.
\newblock {The Great Dimming of Betelgeuse: The photosphere as revealed by
  tomography over the past 15 yr}.
\newblock {\em \aap} {\bf 2024}, {\em 685},~A124,
  \href{http://arxiv.org/abs/2312.02816}{{\normalfont
  [arXiv:astro-ph.SR/2312.02816]}}.
\newblock {\url{https://doi.org/10.1051/0004-6361/202348846}}.

\bibitem[{Kervella} et~al.(2018){Kervella}, {Decin}, {Richards}, {Harper},
  {McDonald}, {O'Gorman}, {Montarg{\`e}s}, {Homan}, and {Ohnaka}]{Kervella2018}
{Kervella}, P.; {Decin}, L.; {Richards}, A.M.S.; {Harper}, G.M.; {McDonald},
  I.; {O'Gorman}, E.; {Montarg{\`e}s}, M.; {Homan}, W.; {Ohnaka}, K.
\newblock {The close circumstellar environment of Betelgeuse. V. Rotation
  velocity and molecular envelope properties from ALMA}.
\newblock {\em \aap} {\bf 2018}, {\em 609},~A67,
  \href{http://arxiv.org/abs/1711.07983}{{\normalfont
  [arXiv:astro-ph.SR/1711.07983]}}.
\newblock {\url{https://doi.org/10.1051/0004-6361/201731761}}.

\bibitem[{Lobel} and {Dupree}(2001)]{2001ApJ...558..815L}
{Lobel}, A.; {Dupree}, A.K.
\newblock {Spatially Resolved STIS Spectroscopy of {\ensuremath{\alpha}}
  Orionis: Evidence for Nonradial Chromospheric Oscillation from Detailed
  Modeling}.
\newblock {\em \apj} {\bf 2001}, {\em 558},~815--829,
  \href{http://arxiv.org/abs/astro-ph/0106548}{{\normalfont
  [arXiv:astro-ph/astro-ph/0106548]}}.
\newblock {\url{https://doi.org/10.1086/322284}}.

\bibitem[{Chiavassa} et~al.(2011){Chiavassa}, {Pasquato}, {Jorissen}, {Sacuto},
  {Babusiaux}, {Freytag}, {Ludwig}, {Cruzal{\`e}bes}, {Rabbia}, {Spang}, and
  {Chesneau}]{2011A&A...528A.120C}
{Chiavassa}, A.; {Pasquato}, E.; {Jorissen}, A.; {Sacuto}, S.; {Babusiaux}, C.;
  {Freytag}, B.; {Ludwig}, H.G.; {Cruzal{\`e}bes}, P.; {Rabbia}, Y.; {Spang},
  A.;  et~al.
\newblock {Radiative hydrodynamic simulations of red supergiant stars. III.
  Spectro-photocentric variability, photometric variability, and consequences
  on Gaia measurements}.
\newblock {\em \aap} {\bf 2011}, {\em 528},~A120,
  \href{http://arxiv.org/abs/1012.5234}{{\normalfont
  [arXiv:astro-ph.SR/1012.5234]}}.
\newblock {\url{https://doi.org/10.1051/0004-6361/201015768}}.

\bibitem[{Chiavassa} et~al.(2022){Chiavassa}, {Kudritzki}, {Davies}, {Freytag},
  and {de Mink}]{2022A&A...661L...1C}
{Chiavassa}, A.; {Kudritzki}, R.; {Davies}, B.; {Freytag}, B.; {de Mink}, S.E.
\newblock {Probing red supergiant dynamics through photo-center displacements
  measured by Gaia}.
\newblock {\em \aap} {\bf 2022}, {\em 661},~L1,
  \href{http://arxiv.org/abs/2205.05156}{{\normalfont
  [arXiv:astro-ph.SR/2205.05156]}}.
\newblock {\url{https://doi.org/10.1051/0004-6361/202243568}}.

\bibitem[{ESA}(1997)]{1997ESASP1200.....E}
{ESA}., Ed.
\newblock {\em {The HIPPARCOS and TYCHO catalogues. Astrometric and photometric
  star catalogues derived from the ESA HIPPARCOS Space Astrometry Mission}},
  Vol. 1200, {\em ESA Special Publication},  1997.

\bibitem[{Perryman} et~al.(1997){Perryman}, {Lindegren}, {Kovalevsky}, {Hoeg},
  {Bastian}, {Bernacca}, {Cr{\'e}z{\'e}}, {Donati}, {Grenon}, {Grewing}, {van
  Leeuwen}, {van der Marel}, {Mignard}, {Murray}, {Le Poole}, {Schrijver},
  {Turon}, {Arenou}, {Froeschl{\'e}}, and {Petersen}]{1997A&A...323L..49P}
{Perryman}, M.A.C.; {Lindegren}, L.; {Kovalevsky}, J.; {Hoeg}, E.; {Bastian},
  U.; {Bernacca}, P.L.; {Cr{\'e}z{\'e}}, M.; {Donati}, F.; {Grenon}, M.;
  {Grewing}, M.;  et~al.
\newblock {The HIPPARCOS Catalogue}.
\newblock {\em \aap} {\bf 1997}, {\em 323},~L49--L52.

\bibitem[{van Leeuwen}(2007)]{2007A&A...474..653V}
{van Leeuwen}, F.
\newblock {Validation of the new Hipparcos reduction}.
\newblock {\em \aap} {\bf 2007}, {\em 474},~653--664,
  \href{http://arxiv.org/abs/0708.1752}{{\normalfont [0708.1752]}}.
\newblock {\url{https://doi.org/10.1051/0004-6361:20078357}}.

\bibitem[Harper et~al.(2017)Harper, Brown, Guinan, O'Gorman, Richards,
  Kervella, and Decin]{2017AJ....154...11H}
Harper, G.M.; Brown, A.; Guinan, E.F.; O'Gorman, E.; Richards, A.M.S.;
  Kervella, P.; Decin, L.
\newblock An Updated 2017 Astrometric Solution for Betelgeuse.
\newblock {\em \aj} {\bf 2017}, {\em 154},~11,
  \href{http://arxiv.org/abs/1706.06020}{{\normalfont
  [arXiv:astro-ph.SR/1706.06020]}}.
\newblock {\url{https://doi.org/10.3847/1538-3881/aa6ff9}}.

\bibitem[{Harper} et~al.(2008){Harper}, {Brown}, and
  {Guinan}]{2008AJ....135.1430H}
{Harper}, G.M.; {Brown}, A.; {Guinan}, E.F.
\newblock {A New VLA-Hipparcos Distance to Betelgeuse and its Implications}.
\newblock {\em \aj} {\bf 2008}, {\em 135},~1430--1440.
\newblock {\url{https://doi.org/10.1088/0004-6256/135/4/1430}}.

\bibitem[{Joyce} et~al.(2020){Joyce}, {Leung}, {Moln{\'a}r}, {Ireland},
  {Kobayashi}, and {Nomoto}]{2020ApJ...902...63J}
{Joyce}, M.; {Leung}, S.C.; {Moln{\'a}r}, L.; {Ireland}, M.; {Kobayashi}, C.;
  {Nomoto}, K.
\newblock {Standing on the Shoulders of Giants: New Mass and Distance Estimates
  for Betelgeuse through Combined Evolutionary, Asteroseismic, and Hydrodynamic
  Simulations with MESA}.
\newblock {\em \apj} {\bf 2020}, {\em 902},~63,
  \href{http://arxiv.org/abs/2006.09837}{{\normalfont
  [arXiv:astro-ph.SR/2006.09837]}}.
\newblock {\url{https://doi.org/10.3847/1538-4357/abb8db}}.

\bibitem[{Dupree} et~al.(2022){Dupree}, {Strassmeier}, {Calderwood}, {Granzer},
  {Weber}, {Kravchenko}, {Matthews}, {Montarg{\`e}s}, {Tappin}, and
  {Thompson}]{2022ApJ...936...18D}
{Dupree}, A.K.; {Strassmeier}, K.G.; {Calderwood}, T.; {Granzer}, T.; {Weber},
  M.; {Kravchenko}, K.; {Matthews}, L.D.; {Montarg{\`e}s}, M.; {Tappin}, J.;
  {Thompson}, W.T.
\newblock {The Great Dimming of Betelgeuse: A Surface Mass Ejection and Its
  Consequences}.
\newblock {\em \apj} {\bf 2022}, {\em 936},~18,
  \href{http://arxiv.org/abs/2208.01676}{{\normalfont
  [arXiv:astro-ph.SR/2208.01676]}}.
\newblock {\url{https://doi.org/10.3847/1538-4357/ac7853}}.

\bibitem[{Levesque} et~al.(2005){Levesque}, {Massey}, {Olsen}, {Plez},
  {Josselin}, {Maeder}, and {Meynet}]{2005ApJ...628..973L}
{Levesque}, E.M.; {Massey}, P.; {Olsen}, K.A.G.; {Plez}, B.; {Josselin}, E.;
  {Maeder}, A.; {Meynet}, G.
\newblock {The Effective Temperature Scale of Galactic Red Supergiants: Cool,
  but Not As Cool As We Thought}.
\newblock {\em \apj} {\bf 2005}, {\em 628},~973--985,
  \href{http://arxiv.org/abs/astro-ph/0504337}{{\normalfont
  [astro-ph/0504337]}}.
\newblock {\url{https://doi.org/10.1086/430901}}.

\bibitem[{Famaey} et~al.(2005){Famaey}, {Jorissen}, {Luri}, {Mayor}, {Udry},
  {Dejonghe}, and {Turon}]{Famaey2005}
{Famaey}, B.; {Jorissen}, A.; {Luri}, X.; {Mayor}, M.; {Udry}, S.; {Dejonghe},
  H.; {Turon}, C.
\newblock Local kinematics of K and M giants from CORAVEL-Hipparco-Tycho-2
  data. Revisiting the concept of superclusters.
\newblock {\em \aap} {\bf 2005}, {\em 430},~165--186,
  \href{http://arxiv.org/abs/astro-ph/0409579}{{\normalfont
  [arXiv:astro-ph/astro-ph/0409579]}}.
\newblock {\url{https://doi.org/10.1051/0004-6361:20041272}}.

\bibitem[{Ma} et~al.(2024){Ma}, {Chiavassa}, {de Mink}, {Valli}, {Justham}, and
  {Freytag}]{2024ApJ...962L..36M}
{Ma}, J.Z.; {Chiavassa}, A.; {de Mink}, S.E.; {Valli}, R.; {Justham}, S.;
  {Freytag}, B.
\newblock {Is Betelgeuse Really Rotating? Synthetic ALMA Observations of
  Large-scale Convection in 3D Simulations of Red Supergiants}.
\newblock {\em \apjl} {\bf 2024}, {\em 962},~L36,
  \href{http://arxiv.org/abs/2311.16885}{{\normalfont
  [arXiv:astro-ph.SR/2311.16885]}}.
\newblock {\url{https://doi.org/10.3847/2041-8213/ad24fd}}.

\bibitem[{De Beck} et~al.(2010){De Beck}, {Decin}, {de Koter}, {Justtanont},
  {Verhoelst}, {Kemper}, and {Menten}]{DeBeck2010}
{De Beck}, E.; {Decin}, L.; {de Koter}, A.; {Justtanont}, K.; {Verhoelst}, T.;
  {Kemper}, F.; {Menten}, K.M.
\newblock {Probing the mass-loss history of AGB and red supergiant stars from
  CO rotational line profiles. II. CO line survey of evolved stars: derivation
  of mass-loss rate formulae}.
\newblock {\em \aap} {\bf 2010}, {\em 523},~A18,
  \href{http://arxiv.org/abs/1008.1083}{{\normalfont
  [arXiv:astro-ph.SR/1008.1083]}}.
\newblock {\url{https://doi.org/10.1051/0004-6361/200913771}}.

\bibitem[{Dolan} et~al.(2016){Dolan}, {Mathews}, {Lam}, {Quynh Lan}, {Herczeg},
  and {Dearborn}]{2016ApJ...819....7D}
{Dolan}, M.M.; {Mathews}, G.J.; {Lam}, D.D.; {Quynh Lan}, N.; {Herczeg}, G.J.;
  {Dearborn}, D.S.P.
\newblock {Evolutionary Tracks for Betelgeuse}.
\newblock {\em \apj} {\bf 2016}, {\em 819},~7.
\newblock {\url{https://doi.org/10.3847/0004-637X/819/1/7}}.

\bibitem[{Saio} et~al.(2023){Saio}, {Nandal}, {Meynet}, and
  {Ekstr{\"o}m}]{2023MNRAS.526.2765S}
{Saio}, H.; {Nandal}, D.; {Meynet}, G.; {Ekstr{\"o}m}, S.
\newblock {The evolutionary stage of Betelgeuse inferred from its pulsation
  periods}.
\newblock {\em \mnras} {\bf 2023}, {\em 526},~2765--2775,
  \href{http://arxiv.org/abs/2306.00287}{{\normalfont
  [arXiv:astro-ph.SR/2306.00287]}}.
\newblock {\url{https://doi.org/10.1093/mnras/stad2949}}.

\bibitem[{Wheeler} et~al.(2017){Wheeler}, {Nance}, {Diaz}, {Smith}, {Hickey},
  {Zhou}, {Koutoulaki}, {Sullivan}, and {Fowler}]{2017MNRAS.465.2654W}
{Wheeler}, J.C.; {Nance}, S.; {Diaz}, M.; {Smith}, S.G.; {Hickey}, J.; {Zhou},
  L.; {Koutoulaki}, M.; {Sullivan}, J.M.; {Fowler}, J.M.
\newblock {The Betelgeuse Project: constraints from rotation}.
\newblock {\em \mnras} {\bf 2017}, {\em 465},~2654--2661,
  \href{http://arxiv.org/abs/1611.08031}{{\normalfont
  [arXiv:astro-ph.SR/1611.08031]}}.
\newblock {\url{https://doi.org/10.1093/mnras/stw2893}}.

\bibitem[{Chatzopoulos} et~al.(2020){Chatzopoulos}, {Frank}, {Marcello}, and
  {Clayton}]{2020ApJ...896...50C}
{Chatzopoulos}, E.; {Frank}, J.; {Marcello}, D.C.; {Clayton}, G.C.
\newblock {Is Betelgeuse the Outcome of a Past Merger?}
\newblock {\em \apj} {\bf 2020}, {\em 896},~50,
  \href{http://arxiv.org/abs/2005.04172}{{\normalfont
  [arXiv:astro-ph.SR/2005.04172]}}.
\newblock {\url{https://doi.org/10.3847/1538-4357/ab91bb}}.

\bibitem[{Sullivan} et~al.(2020){Sullivan}, {Nance}, and
  {Wheeler}]{2020ApJ...905..128S}
{Sullivan}, J.M.; {Nance}, S.; {Wheeler}, J.C.
\newblock {The Betelgeuse Project. III. Merger Characteristics}.
\newblock {\em \apj} {\bf 2020}, {\em 905},~128,
  \href{http://arxiv.org/abs/2010.08880}{{\normalfont
  [arXiv:astro-ph.HE/2010.08880]}}.
\newblock {\url{https://doi.org/10.3847/1538-4357/abc3c9}}.

\bibitem[{Sana} et~al.(2012){Sana}, {de Mink}, {de Koter}, {Langer}, {Evans},
  {Gieles}, {Gosset}, {Izzard}, {Le Bouquin}, and
  {Schneider}]{2012Sci...337..444S}
{Sana}, H.; {de Mink}, S.E.; {de Koter}, A.; {Langer}, N.; {Evans}, C.J.;
  {Gieles}, M.; {Gosset}, E.; {Izzard}, R.G.; {Le Bouquin}, J.B.; {Schneider},
  F.R.N.
\newblock {Binary Interaction Dominates the Evolution of Massive Stars}.
\newblock {\em Science} {\bf 2012}, {\em 337},~444,
  \href{http://arxiv.org/abs/1207.6397}{{\normalfont
  [arXiv:astro-ph.SR/1207.6397]}}.
\newblock {\url{https://doi.org/10.1126/science.1223344}}.

\bibitem[{Bordier} et~al.(2022){Bordier}, {Frost}, {Sana}, {Reggiani},
  {M{\'e}rand}, {Rainot}, {Ram{\'\i}rez-Tannus}, and {de
  Wit}]{2022A&A...663A..26B}
{Bordier}, E.; {Frost}, A.J.; {Sana}, H.; {Reggiani}, M.; {M{\'e}rand}, A.;
  {Rainot}, A.; {Ram{\'\i}rez-Tannus}, M.C.; {de Wit}, W.J.
\newblock {The origin of close massive binaries in the M17 star-forming
  region}.
\newblock {\em \aap} {\bf 2022}, {\em 663},~A26,
  \href{http://arxiv.org/abs/2203.05036}{{\normalfont
  [arXiv:astro-ph.SR/2203.05036]}}.
\newblock {\url{https://doi.org/10.1051/0004-6361/202141849}}.

\bibitem[{Goldberg} et~al.(2020){Goldberg}, {Bauer}, and
  {Howell}]{2020RNAAS...4...35G}
{Goldberg}, J.A.; {Bauer}, E.B.; {Howell}, D.A.
\newblock {Apparent Magnitude of Betelgeuse as a Type IIP Supernova}.
\newblock {\em Research Notes of the American Astronomical Society} {\bf 2020},
  {\em 4},~35.
\newblock {\url{https://doi.org/10.3847/2515-5172/ab7c68}}.

\bibitem[{Moln{\'a}r} et~al.(2023){Moln{\'a}r}, {Joyce}, and
  {Leung}]{2023RNAAS...7..119M}
{Moln{\'a}r}, L.; {Joyce}, M.; {Leung}, S.C.
\newblock {Comment on the Feasibility of Carbon Burning in Betelgeuse}.
\newblock {\em Research Notes of the American Astronomical Society} {\bf 2023},
  {\em 7},~119,  \href{http://arxiv.org/abs/2306.05600}{{\normalfont
  [arXiv:astro-ph.SR/2306.05600]}}.
\newblock {\url{https://doi.org/10.3847/2515-5172/acdb7a}}.

\bibitem[{Montarg{\`e}s} et~al.(2014){Montarg{\`e}s}, {Kervella}, {Perrin},
  {Ohnaka}, {Chiavassa}, {Ridgway}, and {Lacour}]{2014A&A...572A..17M}
{Montarg{\`e}s}, M.; {Kervella}, P.; {Perrin}, G.; {Ohnaka}, K.; {Chiavassa},
  A.; {Ridgway}, S.T.; {Lacour}, S., M.
\newblock {Properties of the CO and H$_{2}$O MOLsphere of the red supergiant
  Betelgeuse from VLTI/AMBER observations}.
\newblock {\em \aap} {\bf 2014}, {\em 572},~A17,
  \href{http://arxiv.org/abs/1408.2994}{{\normalfont
  [arXiv:astro-ph.SR/1408.2994]}}.
\newblock {\url{https://doi.org/10.1051/0004-6361/201423538}}.

\bibitem[{Ekstr{\"o}m} et~al.(2012){Ekstr{\"o}m}, {Georgy}, {Eggenberger},
  {Meynet}, {Mowlavi}, {Wyttenbach}, {Granada}, {Decressin}, {Hirschi},
  {Frischknecht}, {Charbonnel}, and {Maeder}]{2012A&A...537A.146E}
{Ekstr{\"o}m}, S.; {Georgy}, C.; {Eggenberger}, P.; {Meynet}, G.; {Mowlavi},
  N.; {Wyttenbach}, A.; {Granada}, A.; {Decressin}, T.; {Hirschi}, R.;
  {Frischknecht}, U.;  et~al.
\newblock {Grids of stellar models with rotation. I. Models from 0.8 to 120
  M$_{\&sun;}$ at solar metallicity (Z = 0.014)}.
\newblock {\em \aap} {\bf 2012}, {\em 537},~A146,
  \href{http://arxiv.org/abs/1110.5049}{{\normalfont
  [arXiv:astro-ph.SR/1110.5049]}}.
\newblock {\url{https://doi.org/10.1051/0004-6361/201117751}}.

\bibitem[{Humphreys} et~al.(2021){Humphreys}, {Davidson}, {Richards}, {Ziurys},
  {Jones}, and {Ishibashi}]{2021AJ....161...98H}
{Humphreys}, R.M.; {Davidson}, K.; {Richards}, A.M.S.; {Ziurys}, L.M.; {Jones},
  T.J.; {Ishibashi}, K.
\newblock {The Mass-loss History of the Red Hypergiant VY CMa}.
\newblock {\em \aj} {\bf 2021}, {\em 161},~98,
  \href{http://arxiv.org/abs/2012.08550}{{\normalfont
  [arXiv:astro-ph.SR/2012.08550]}}.
\newblock {\url{https://doi.org/10.3847/1538-3881/abd316}}.

\bibitem[{Young} et~al.(1993){Young}, {Phillips}, and
  {Knapp}]{1993ApJS...86..517Y}
{Young}, K.; {Phillips}, T.G.; {Knapp}, G.R.
\newblock {Circumstellar Shells Resolved in the IRAS Survey Data. I. Data
  Processing Procedure, Results, and Confidence Tests}.
\newblock {\em \apjs} {\bf 1993}, {\em 86},~517.
\newblock {\url{https://doi.org/10.1086/191789}}.

\bibitem[{Noriega-Crespo} et~al.(1997){Noriega-Crespo}, {van Buren}, {Cao}, and
  {Dgani}]{1997AJ....114..837N}
{Noriega-Crespo}, A.; {van Buren}, D.; {Cao}, Y.; {Dgani}, R.
\newblock {A Parsec-Size Bow Shock around Betelgeuse}.
\newblock {\em \aj} {\bf 1997}, {\em 114},~837--840.
\newblock {\url{https://doi.org/10.1086/118517}}.

\bibitem[{Le Bertre} et~al.(2012){Le Bertre}, {Matthews}, {G{\'e}rard}, and
  {Libert}]{2012MNRAS.422.3433L}
{Le Bertre}, T.; {Matthews}, L.D.; {G{\'e}rard}, E.; {Libert}, Y.
\newblock {Discovery of a detached H I gas shell surrounding {$\alpha$}
  Orionis}.
\newblock {\em \mnras} {\bf 2012}, {\em 422},~3433--3443,
  \href{http://arxiv.org/abs/1203.0255}{{\normalfont
  [arXiv:astro-ph.SR/1203.0255]}}.
\newblock {\url{https://doi.org/10.1111/j.1365-2966.2012.20853.x}}.

\bibitem[{Wood} et~al.(2004){Wood}, {Olivier}, and
  {Kawaler}]{2004ASPC..310..322W}
{Wood}, P.R.; {Olivier}, A.E.; {Kawaler}, S.D.
\newblock {The long secondary periods in semi-regular variables}.
\newblock In Proceedings of the IAU Colloq. 193: Variable Stars in the Local
  Group; {Kurtz}, D.W.; {Pollard}, K.R., Eds.,  2004, Vol. 310, {\em
  Astronomical Society of the Pacific Conference Series}, p. 322.

\bibitem[{Goldberg} et~al.(2024){Goldberg}, {Joyce}, and
  {Moln{\'a}r}]{2024ApJ...977...35G}
{Goldberg}, J.A.; {Joyce}, M.; {Moln{\'a}r}, L.
\newblock {A Buddy for Betelgeuse: Binarity as the Origin of the Long Secondary
  Period in {\ensuremath{\alpha}} Orionis}.
\newblock {\em \apj} {\bf 2024}, {\em 977},~35,
  \href{http://arxiv.org/abs/2408.09089}{{\normalfont
  [arXiv:astro-ph.SR/2408.09089]}}.
\newblock {\url{https://doi.org/10.3847/1538-4357/ad87f4}}.

\bibitem[{Granzer} et~al.(2022){Granzer}, {Weber}, {Strassmeier}, and
  {Dupree}]{2022csss.confE.185G}
{Granzer}, T.; {Weber}, M.; {Strassmeier}, K.G.; {Dupree}, A.
\newblock {Betelgeuse: Long Secondary Period, a Fundamental Mode and
  Overtones}.
\newblock In Proceedings of the The 21st Cambridge Workshop on Cool Stars,
  Stellar Systems, and the Sun,  2022, Cambridge Workshop on Cool Stars,
  Stellar Systems, and the Sun, p. 185.
\newblock {\url{https://doi.org/10.5281/zenodo.7589936}}.

\bibitem[{MacLeod} et~al.(2023){MacLeod}, {Antoni}, {Huang}, {Dupree}, and
  {Loeb}]{2023ApJ...956...27M}
{MacLeod}, M.; {Antoni}, A.; {Huang}, C.D.; {Dupree}, A.; {Loeb}, A.
\newblock {Left Ringing: Betelgeuse Illuminates the Connection between
  Convective Outbursts, Mode Switching, and Mass Ejection in Red Supergiants}.
\newblock {\em \apj} {\bf 2023}, {\em 956},~27,
  \href{http://arxiv.org/abs/2305.09732}{{\normalfont
  [arXiv:astro-ph.SR/2305.09732]}}.
\newblock {\url{https://doi.org/10.3847/1538-4357/aced4b}}.

\bibitem[{Granzer} et~al.(2021){Granzer}, {Weber}, {Strassmeier}, and
  {Dupree}]{2021csss.confE..41G}
{Granzer}, T.; {Weber}, M.; {Strassmeier}, K.G.; {Dupree}, A.
\newblock {The Curious Case of Betelgeuse}.
\newblock In Proceedings of the The 20.5th Cambridge Workshop on Cool Stars,
  Stellar Systems, and the Sun (CS20.5),  2021, Cambridge Workshop on Cool
  Stars, Stellar Systems, and the Sun, p.~41.
\newblock {\url{https://doi.org/10.5281/zenodo.4561732}}.

\bibitem[{Taniguchi} et~al.(2021){Taniguchi}, {Matsunaga}, {Jian}, {Kobayashi},
  {Fukue}, {Hamano}, {Ikeda}, {Kawakita}, {Kondo}, {Otsubo}, {Sameshima},
  {Takenaka}, and {Yasui}]{2021MNRAS.502.4210T}
{Taniguchi}, D.; {Matsunaga}, N.; {Jian}, M.; {Kobayashi}, N.; {Fukue}, K.;
  {Hamano}, S.; {Ikeda}, Y.; {Kawakita}, H.; {Kondo}, S.; {Otsubo}, S.;  et~al.
\newblock {Effective temperatures of red supergiants estimated from line-depth
  ratios of iron lines in the YJ bands, 0.97-1.32{\ensuremath{\mu}}m}.
\newblock {\em \mnras} {\bf 2021}, {\em 502},~4210--4226,
  \href{http://arxiv.org/abs/2012.07856}{{\normalfont
  [arXiv:astro-ph.SR/2012.07856]}}.
\newblock {\url{https://doi.org/10.1093/mnras/staa3855}}.

\bibitem[{Taniguchi} et~al.(2022){Taniguchi}, {Yamazaki}, and
  {Uno}]{2022NatAs...6..930T}
{Taniguchi}, D.; {Yamazaki}, K.; {Uno}, S.
\newblock {The Great Dimming of Betelgeuse seen by the Himawari-8
  meteorological satellite}.
\newblock {\em Nature Astronomy} {\bf 2022}, {\em 6},~930--935,
  \href{http://arxiv.org/abs/2205.14165}{{\normalfont
  [arXiv:astro-ph.SR/2205.14165]}}.
\newblock {\url{https://doi.org/10.1038/s41550-022-01680-5}}.

\bibitem[{Matthews} and {Dupree}(2022)]{2022ApJ...934..131M}
{Matthews}, L.D.; {Dupree}, A.K.
\newblock {Spatially Resolved Observations of Betelgeuse at
  {\ensuremath{\lambda}}7 mm and {\ensuremath{\lambda}}1.3 cm Just prior to the
  Great Dimming}.
\newblock {\em \apj} {\bf 2022}, {\em 934},~131,
  \href{http://arxiv.org/abs/2206.04144}{{\normalfont
  [arXiv:astro-ph.SR/2206.04144]}}.
\newblock {\url{https://doi.org/10.3847/1538-4357/ac7726}}.

\bibitem[{Alexeeva} et~al.(2021){Alexeeva}, {Zhao}, {Gao}, {Du}, {Li}, {Li},
  and {Hu}]{2021NatCo..12.4719A}
{Alexeeva}, S.; {Zhao}, G.; {Gao}, D.Y.; {Du}, J.; {Li}, A.; {Li}, K.; {Hu}, S.
\newblock {Spectroscopic evidence for a large spot on the dimming Betelgeuse}.
\newblock {\em Nature Communications} {\bf 2021}, {\em 12},~4719,
  \href{http://arxiv.org/abs/2108.03472}{{\normalfont
  [arXiv:astro-ph.SR/2108.03472]}}.
\newblock {\url{https://doi.org/10.1038/s41467-021-25018-3}}.

\bibitem[{Guinan} et~al.(2019){Guinan}, {Wasatonic}, and
  {Calderwood}]{2019ATel13341....1G}
{Guinan}, E.F.; {Wasatonic}, R.J.; {Calderwood}, T.J.
\newblock {The Fainting of the Nearby Red Supergiant Betelgeuse}.
\newblock {\em The Astronomer's Telegram} {\bf 2019}, {\em 13341},~1.

\bibitem[Levesque and Massey(2020)]{2020ApJ...891L..37L}
Levesque, E.M.; Massey, P.
\newblock Betelgeuse Just Is Not That Cool: Effective Temperature Alone Cannot
  Explain the Recent Dimming of Betelgeuse.
\newblock {\em \apjl} {\bf 2020}, {\em 891},~L37,
  \href{http://arxiv.org/abs/2002.10463}{{\normalfont
  [arXiv:astro-ph.SR/2002.10463]}}.
\newblock {\url{https://doi.org/10.3847/2041-8213/ab7935}}.

\bibitem[Cotton et~al.(2020)Cotton, Bailey, De~Horta, Norris, and
  Lomax]{2020RNAAS...4...39C}
Cotton, D.V.; Bailey, J.; De~Horta, A.Y.; Norris, B.R.M.; Lomax, J.R.
\newblock Multi-band Aperture Polarimetry of Betelgeuse during the 2019-20
  Dimming.
\newblock {\em Research Notes of the American Astronomical Society} {\bf 2020},
  {\em 4},~39.
\newblock {\url{https://doi.org/10.3847/2515-5172/ab7f2f}}.

\bibitem[Safonov et~al.(2020)Safonov, Dodin, Burlak, Goliguzova, Fedoteva,
  Zheltoukhov, Lamzin, Strakhov, and Voziakova]{2020arXiv200505215S}
Safonov, B.; Dodin, A.; Burlak, M.; Goliguzova, M.; Fedoteva, A.; Zheltoukhov,
  S.; Lamzin, S.; Strakhov, I.; Voziakova, O.
\newblock Differential Speckle Polarimetry of Betelgeuse in 2019-2020: the rise
  is different from the fall.
\newblock {\em arXiv e-prints} {\bf 2020}, p. arXiv:2005.05215,
  \href{http://arxiv.org/abs/2005.05215}{{\normalfont
  [arXiv:astro-ph.SR/2005.05215]}}.
\newblock {\url{https://doi.org/10.48550/arXiv.2005.05215}}.

\bibitem[Lan{\c{c}}on et~al.(2007)Lan{\c{c}}on, Hauschildt, Ladjal, and
  Mouhcine]{2007A&A...468..205L}
Lan{\c{c}}on, A.; Hauschildt, P.H.; Ladjal, D.; Mouhcine, M.
\newblock Near-IR spectra of red supergiants and giants. I. Models with solar
  and with mixing-induced surface abundance ratios.
\newblock {\em \aap} {\bf 2007}, {\em 468},~205--220,
  \href{http://arxiv.org/abs/0704.2120}{{\normalfont
  [arXiv:astro-ph/0704.2120]}}.
\newblock {\url{https://doi.org/10.1051/0004-6361:20065824}}.

\bibitem[Dullemond et~al.(2012)Dullemond, Juhasz, Pohl, Sereshti, Shetty,
  Peters, Commercon, and Flock]{2012ascl.soft02015D}
Dullemond, C.P.; Juhasz, A.; Pohl, A.; Sereshti, F.; Shetty, R.; Peters, T.;
  Commercon, B.; Flock, M.
\newblock RADMC-3D: A multi-purpose radiative transfer tool.
\newblock Astrophysics Source Code Library, record ascl:1202.015,  2012.

\bibitem[{Kashyap} et~al.(2020){Kashyap}, {Drake}, and
  {Patnaude}]{2020ATel13501....1K}
{Kashyap}, V.L.; {Drake}, J.J.; {Patnaude}, D.
\newblock {Non-detection of Betelgeuse in X-rays}.
\newblock {\em The Astronomer's Telegram} {\bf 2020}, {\em 13501},~1.

\bibitem[{Freytag} et~al.(2024){Freytag}, {H{\"o}fner}, {Aringer}, and
  {Chiavassa}]{2024A&A...692A.223F}
{Freytag}, B.; {H{\"o}fner}, S.; {Aringer}, B.; {Chiavassa}, A.
\newblock {Dimming events of evolved stars due to clouds of molecular gas:
  Scenarios based on 3D radiation-hydrodynamics simulations with CO5BOLD}.
\newblock {\em \aap} {\bf 2024}, {\em 692},~A223,
  \href{http://arxiv.org/abs/2411.17561}{{\normalfont
  [arXiv:astro-ph.SR/2411.17561]}}.
\newblock {\url{https://doi.org/10.1051/0004-6361/202450829}}.

\bibitem[{Tappin} et~al.(2015){Tappin}, {Eyles}, and
  {Davies}]{2015SoPh..290.2143T}
{Tappin}, S.J.; {Eyles}, C.J.; {Davies}, J.A.
\newblock {Determination of the Photometric Calibration and Large-Scale
  Flatfield of the STEREO Heliospheric Imagers: II. HI-2}.
\newblock {\em \solphys} {\bf 2015}, {\em 290},~2143--2170.
\newblock {\url{https://doi.org/10.1007/s11207-015-0737-5}}.

\bibitem[{Dupree} et~al.(2020){Dupree}, {Guinan}, {Thompson}, and
  {STEREO/SECCHI/HI Consortium}]{2020ATel13901....1D}
{Dupree}, A.; {Guinan}, E.; {Thompson}, W.T.; {STEREO/SECCHI/HI Consortium}.
\newblock {Photometry of Betelgeuse with the STEREO Mission While in the Glare
  of the Sun from Earth}.
\newblock {\em The Astronomer's Telegram} {\bf 2020}, {\em 13901},~1.

\bibitem[{Soszy{\'n}ski} et~al.(2021){Soszy{\'n}ski}, {Olechowska},
  {Ratajczak}, {Iwanek}, {Skowron}, {Mr{\'o}z}, {Pietrukowicz}, {Udalski},
  {Szyma{\'n}ski}, {Skowron}, {Gromadzki}, {Poleski}, {Koz{\l}owski}, {Wrona},
  {Ulaczyk}, and {Rybicki}]{2021ApJ...911L..22S}
{Soszy{\'n}ski}, I.; {Olechowska}, A.; {Ratajczak}, M.; {Iwanek}, P.;
  {Skowron}, D.M.; {Mr{\'o}z}, P.; {Pietrukowicz}, P.; {Udalski}, A.;
  {Szyma{\'n}ski}, M.K.; {Skowron}, J.;  et~al.
\newblock {Binarity as the Origin of Long Secondary Periods in Red Giant
  Stars}.
\newblock {\em \apjl} {\bf 2021}, {\em 911},~L22,
  \href{http://arxiv.org/abs/2103.12748}{{\normalfont
  [arXiv:astro-ph.SR/2103.12748]}}.
\newblock {\url{https://doi.org/10.3847/2041-8213/abf3c9}}.

\bibitem[{MacLeod} et~al.(2025){MacLeod}, {Blunt}, {De Rosa}, {Dupree},
  {Granzer}, {Harper}, {Huang}, {Leiner}, {Loeb}, {Nielsen}, {Strassmeier},
  {Wang}, and {Weber}]{2025ApJ...978...50M}
{MacLeod}, M.; {Blunt}, S.; {De Rosa}, R.J.; {Dupree}, A.K.; {Granzer}, T.;
  {Harper}, G.M.; {Huang}, C.D.; {Leiner}, E.M.; {Loeb}, A.; {Nielsen}, E.L.;
  et~al.
\newblock {Radial Velocity and Astrometric Evidence for a Close Companion to
  Betelgeuse}.
\newblock {\em \apj} {\bf 2025}, {\em 978},~50,
  \href{http://arxiv.org/abs/2409.11332}{{\normalfont
  [arXiv:astro-ph.SR/2409.11332]}}.
\newblock {\url{https://doi.org/10.3847/1538-4357/ad93c8}}.

\bibitem[{Brandt} et~al.(1995){Brandt}, {Heap}, {Beaver}, {Boggess},
  {Carpenter}, {Ebbets}, {Hutchings}, {Jura}, {Leckrone}, {Linsky}, {Maran},
  {Savage}, {Smith}, {Trafton}, {Walter}, {Weymann}, {Snow}, {Randall}, {Ake},
  {Robinson}, and {Wahlgren}]{1995AJ....109.2706B}
{Brandt}, J.C.; {Heap}, S.R.; {Beaver}, E.A.; {Boggess}, A.; {Carpenter}, K.G.;
  {Ebbets}, D.C.; {Hutchings}, J.B.; {Jura}, M.; {Leckrone}, D.S.; {Linsky},
  J.L.;  et~al.
\newblock {An Atlas of Alpha Orionis Obtained with the Goddard High Resolution
  Spectrograph on the Hubble Space Telescope}.
\newblock {\em \aj} {\bf 1995}, {\em 109},~2706.
\newblock {\url{https://doi.org/10.1086/117484}}.

\bibitem[{Dupree} et~al.(2005){Dupree}, {Lobel}, {Young}, {Ake}, {Linsky}, and
  {Redfield}]{2005ApJ...622..629D}
{Dupree}, A.K.; {Lobel}, A.; {Young}, P.R.; {Ake}, T.B.; {Linsky}, J.L.;
  {Redfield}, S.
\newblock {A Far-Ultraviolet Spectroscopic Survey of Luminous Cool Stars}.
\newblock {\em \apj} {\bf 2005}, {\em 622},~629--652,
  \href{http://arxiv.org/abs/astro-ph/0412539}{{\normalfont
  [arXiv:astro-ph/astro-ph/0412539]}}.
\newblock {\url{https://doi.org/10.1086/428111}}.

\bibitem[{Ortiz} et~al.(2020){Ortiz}, {Santos-Sanz}, {Sicardy},
  {Benedetti-Rossi}, {Duffard}, {Morales}, {Braga-Ribas},
  {Fern{\'a}ndez-Valenzuela}, {Nascimbeni}, {Nardiello}, {Carbognani}, {Buzzi},
  {Aletti}, {Bacci}, {Maestripieri}, {Mazzei}, {Mikuz}, {Skvarc}, {Ciabattari},
  {Lavalade}, {Scarfi}, {Mari}, {Conjat}, {Sposetti}, {Bachini}, {Succi},
  {Mancini}, {Alighieri}, {Dal Canto}, {Masucci}, {Vara-Lubiano},
  {Guti{\'e}rrez}, {Desmars}, {Lecacheux}, {Vieira-Martins}, {Camargo},
  {Assafin}, {Colas}, {Beisker}, {Behrend}, {Mueller}, {Meza}, {Gomes-Junior},
  {Roques}, {Vachier}, {Mottola}, {Hellmich}, {Campo Bagatin},
  {Alvarez-Candal}, {Cikota}, {Cikota}, {Christille}, {P{\'a}l}, {Kiss},
  {Pribulla}, {Kom{\v{z}}{\'\i}k}, {Madiedo}, {Charmandaris}, {Alikakos},
  {Szak{\'a}ts}, {Farkas-Tak{\'a}cs}, {Varga-Vereb{\'e}lyi}, {Marton},
  {Marciniak}, {Bartczak}, {Butkiewicz-Ba{\c{k}}}, {Dudzi{\'n}ski},
  {Al{\'\i}-Lagoa}, {Gazeas}, {Paschalis}, {Tsamis}, {Guirado}, {Peris},
  {Iglesias-Marzoa}, {Schnabel}, {Manzano}, {Navarro}, {Perell{\'o}},
  {Vecchione}, {Noschese}, and {Morrone}]{2020A&A...639A.134O}
{Ortiz}, J.L.; {Santos-Sanz}, P.; {Sicardy}, B.; {Benedetti-Rossi}, G.;
  {Duffard}, R.; {Morales}, N.; {Braga-Ribas}, F.; {Fern{\'a}ndez-Valenzuela},
  E.; {Nascimbeni}, V.; {Nardiello}, D.;  et~al.
\newblock {The large trans-Neptunian object 2002 TC$_{302}$ from combined
  stellar occultation, photometry, and astrometry data}.
\newblock {\em \aap} {\bf 2020}, {\em 639},~A134,
  \href{http://arxiv.org/abs/2005.08881}{{\normalfont
  [arXiv:astro-ph.EP/2005.08881]}}.
\newblock {\url{https://doi.org/10.1051/0004-6361/202038046}}.

\bibitem[{Vara-Lubiano} et~al.(2022){Vara-Lubiano}, {Benedetti-Rossi},
  {Santos-Sanz}, {Ortiz}, {Sicardy}, {Popescu}, {Morales}, {Rommel}, {Morgado},
  {Pereira}, {{\'A}lvarez-Candal}, {Fern{\'a}ndez-Valenzuela}, {Souami},
  {Ilic}, {Vince}, {Bachev}, {Semkov}, {Nedelcu}, {{\c{S}}onka}, {Hudin},
  {Boaca}, {Inceu}, {Curelaru}, {Gherase}, {Turcu}, {Moldovan}, {Mircea},
  {Predatu}, {Teodorescu}, {Stoian}, {Juravle}, {Braga-Ribas}, {Desmars},
  {Duffard}, {Lecacheux}, {Camargo}, {Assafin}, {Vieira-Martins}, {Pribulla},
  {Hus{\'a}rik}, {Sivani{\v{c}}}, {Pal}, {Szakats}, {Kiss}, {Alonso-Santiago},
  {Frasca}, {Szab{\'o}}, {Derekas}, {Szigeti}, {Drozdz}, {Ogloza},
  {Skvar{\v{c}}}, {Ciabattari}, {Delincak}, {Di Marcantonio}, {Iafrate},
  {Coretti}, {Baldini}, {Baruffetti}, {Kl{\"o}s}, {Dumitrescu}, {Miku{\v{z}}},
  and {Mohar}]{2022A&A...663A.121V}
{Vara-Lubiano}, M.; {Benedetti-Rossi}, G.; {Santos-Sanz}, P.; {Ortiz}, J.L.;
  {Sicardy}, B.; {Popescu}, M.; {Morales}, N.; {Rommel}, F.L.; {Morgado}, B.;
  {Pereira}, C.L.;  et~al.
\newblock {The multichord stellar occultation on 2019 October 22 by the
  trans-Neptunian object (84922) 2003 VS$_{2}$}.
\newblock {\em \aap} {\bf 2022}, {\em 663},~A121,
  \href{http://arxiv.org/abs/2205.12878}{{\normalfont
  [arXiv:astro-ph.EP/2205.12878]}}.
\newblock {\url{https://doi.org/10.1051/0004-6361/202141842}}.

\bibitem[{Morgado} et~al.(2022){Morgado}, {Bruno}, {Gomes-J{\'u}nior},
  {Pagano}, {Sicardy}, {Fortier}, {Desmars}, {Maxted}, {Braga-Ribas}, {Queloz},
  {Sousa}, {Ortiz}, {Brandeker}, {Collier Cameron}, {Pereira}, {Flor{\'e}n},
  {Hara}, {Souami}, {Isaak}, {Olofsson}, {Santos-Sanz}, {Wilson}, {Broughton},
  {Alibert}, {Alonso}, {Anglada}, {B{\'a}rczy}, {Barrado}, {Barros},
  {Baumjohann}, {Beck}, {Beck}, {Benz}, {Billot}, {Bonfils}, {Broeg},
  {Cabrera}, {Charnoz}, {Csizmadia}, {Davies}, {Deleuil}, {Delrez},
  {Demangeon}, {Demory}, {Ehrenreich}, {Erikson}, {Fossati}, {Fridlund},
  {Gandolfi}, {Gillon}, {G{\"u}del}, {Heng}, {Hoyer}, {Kiss}, {Laskar},
  {Lecavelier des Etangs}, {Lendl}, {Lovis}, {Magrin}, {Marafatto},
  {Nascimbeni}, {Ottensamer}, {Pall{\'e}}, {Peter}, {Piazza}, {Piotto},
  {Pollacco}, {Ragazzoni}, {Rando}, {Ratti}, {Rauer}, {Reimers}, {Ribas},
  {Santos}, {Scandariato}, {S{\'e}gransan}, {Simon}, {Smith}, {Steller},
  {Szab{\'o}}, {Thomas}, {Udry}, {Van Grootel}, {Walton}, and
  {Westerdorff}]{2022A&A...664L..15M}
{Morgado}, B.E.; {Bruno}, G.; {Gomes-J{\'u}nior}, A.R.; {Pagano}, I.;
  {Sicardy}, B.; {Fortier}, A.; {Desmars}, J.; {Maxted}, P.F.L.; {Braga-Ribas},
  F.; {Queloz}, D.;  et~al.
\newblock {A stellar occultation by the transneptunian object (50000) Quaoar
  observed by CHEOPS}.
\newblock {\em \aap} {\bf 2022}, {\em 664},~L15,
  \href{http://arxiv.org/abs/2208.06204}{{\normalfont
  [arXiv:astro-ph.EP/2208.06204]}}.
\newblock {\url{https://doi.org/10.1051/0004-6361/202244221}}.

\bibitem[{Prada} et~al.(2024){Prada}, {Gomez-Merchan}, {P{\'e}rez},
  {Betancort-Rijo}, {Le{\~n}ero-Bardallo}, {Rodr{\'\i}guez-V{\'a}zquez},
  {Glez-de-Rivera}, {D{\'\i}az-L{\'o}pez}, and {de Elias
  Cantalapiedra}]{2024arXiv240614704P}
{Prada}, F.; {Gomez-Merchan}, R.; {P{\'e}rez}, E.; {Betancort-Rijo}, J.E.;
  {Le{\~n}ero-Bardallo}, J.A.; {Rodr{\'\i}guez-V{\'a}zquez}, {\'A}.;
  {Glez-de-Rivera}, G.; {D{\'\i}az-L{\'o}pez}, S.; {de Elias Cantalapiedra}, J.
\newblock {Single-photon gig in Betelgeuse's occultation}.
\newblock {\em arXiv e-prints} {\bf 2024}, p. arXiv:2406.14704,
  \href{http://arxiv.org/abs/2406.14704}{{\normalfont
  [arXiv:astro-ph.IM/2406.14704]}}.
\newblock {\url{https://doi.org/10.48550/arXiv.2406.14704}}.

\bibitem[{Humphreys} et~al.(2019){Humphreys}, {Ziurys}, {Bernal}, {Gordon},
  {Helton}, {Ishibashi}, {Jones}, {Richards}, and
  {Vlemmings}]{2019ApJ...874L..26H}
{Humphreys}, R.M.; {Ziurys}, L.M.; {Bernal}, J.J.; {Gordon}, M.S.; {Helton},
  L.A.; {Ishibashi}, K.; {Jones}, T.J.; {Richards}, A.M.S.; {Vlemmings}, W.
\newblock {The Unexpected Spectrum of the Innermost Ejecta of the Red
  Hypergiant VY CMa}.
\newblock {\em \apjl} {\bf 2019}, {\em 874},~L26,
  \href{http://arxiv.org/abs/1903.08744}{{\normalfont
  [arXiv:astro-ph.SR/1903.08744]}}.
\newblock {\url{https://doi.org/10.3847/2041-8213/ab11e5}}.

\bibitem[{Humphreys} and {Jones}(2022)]{2022AJ....163..103H}
{Humphreys}, R.M.; {Jones}, T.J.
\newblock {Episodic Gaseous Outflows and Mass Loss from Red Supergiants}.
\newblock {\em \aj} {\bf 2022}, {\em 163},~103,
  \href{http://arxiv.org/abs/2201.07818}{{\normalfont
  [arXiv:astro-ph.SR/2201.07818]}}.
\newblock {\url{https://doi.org/10.3847/1538-3881/ac46ff}}.

\bibitem[{Jencson} et~al.(2022){Jencson}, {Sand}, {Andrews}, {Smith},
  {Pearson}, {Strader}, {Valenti}, {Beasor}, and
  {Rothberg}]{2022ApJ...930...81J}
{Jencson}, J.E.; {Sand}, D.J.; {Andrews}, J.E.; {Smith}, N.; {Pearson}, J.;
  {Strader}, J.; {Valenti}, S.; {Beasor}, E.R.; {Rothberg}, B.
\newblock {An Exceptional Dimming Event for a Massive, Cool Supergiant in M51}.
\newblock {\em \apj} {\bf 2022}, {\em 930},~81,
  \href{http://arxiv.org/abs/2110.11376}{{\normalfont
  [arXiv:astro-ph.SR/2110.11376]}}.
\newblock {\url{https://doi.org/10.3847/1538-4357/ac626c}}.

\bibitem[{Anugu} et~al.(2023){Anugu}, {Baron}, {Gies}, {Lanthermann},
  {Schaefer}, {Shepard}, {Brummelaar}, {Monnier}, {Kraus}, {Le Bouquin},
  {Davies}, {Ennis}, {Gardner}, {Labdon}, {Roettenbacher}, {Setterholm},
  {Vollmann}, and {Sigismondi}]{2023AJ....166...78A}
{Anugu}, N.; {Baron}, F.; {Gies}, D.R.; {Lanthermann}, C.; {Schaefer}, G.H.;
  {Shepard}, K.A.; {Brummelaar}, T.t.; {Monnier}, J.D.; {Kraus}, S.; {Le
  Bouquin}, J.B.;  et~al.
\newblock {The Great Dimming of the Hypergiant Star RW Cephei: CHARA Array
  Images and Spectral Analysis}.
\newblock {\em \aj} {\bf 2023}, {\em 166},~78,
  \href{http://arxiv.org/abs/2307.04926}{{\normalfont
  [arXiv:astro-ph.SR/2307.04926]}}.
\newblock {\url{https://doi.org/10.3847/1538-3881/ace59d}}.

\bibitem[{Anugu} et~al.(2024){Anugu}, {Gies}, {Roettenbacher}, {Monnier},
  {Montarg{\'e}s}, {M{\'e}rand}, {Baron}, {Schaefer}, {Shepard}, {Kraus},
  {Anderson}, {Codron}, {Gardner}, {Gutierrez}, {K{\"o}hler}, {Kubiak},
  {Lanthermann}, {Majoinen}, {Scott}, and {Vollmann}]{2024ApJ...973L...5A}
{Anugu}, N.; {Gies}, D.R.; {Roettenbacher}, R.M.; {Monnier}, J.D.;
  {Montarg{\'e}s}, M.; {M{\'e}rand}, A.; {Baron}, F.; {Schaefer}, G.H.;
  {Shepard}, K.A.; {Kraus}, S.;  et~al.
\newblock {Time Evolution Images of the Hypergiant RW Cephei during the
  Rebrightening Phase Following the Great Dimming}.
\newblock {\em \apjl} {\bf 2024}, {\em 973},~L5,
  \href{http://arxiv.org/abs/2408.11906}{{\normalfont
  [arXiv:astro-ph.SR/2408.11906]}}.
\newblock {\url{https://doi.org/10.3847/2041-8213/ad736c}}.

\bibitem[{Munoz-Sanchez} et~al.(2024){Munoz-Sanchez}, {de Wit}, {Bonanos},
  {Antoniadis}, {Boutsia}, {Boumis}, {Christodoulou}, {Kalitsounaki}, and
  {Udalski}]{2024A&A...690A..99M}
{Munoz-Sanchez}, G.; {de Wit}, S.; {Bonanos}, A.Z.; {Antoniadis}, K.;
  {Boutsia}, K.; {Boumis}, P.; {Christodoulou}, E.; {Kalitsounaki}, M.;
  {Udalski}, A.
\newblock {Episodic mass loss in the very luminous red supergiant [W60] B90 in
  the Large Magellanic Cloud}.
\newblock {\em \aap} {\bf 2024}, {\em 690},~A99,
  \href{http://arxiv.org/abs/2405.11019}{{\normalfont
  [arXiv:astro-ph.SR/2405.11019]}}.
\newblock {\url{https://doi.org/10.1051/0004-6361/202450737}}.

\end{thebibliography}
\end{document}